\documentclass{article}

\usepackage{microtype}
\usepackage{graphicx}
\usepackage{subfigure}
\usepackage{booktabs} 

\usepackage{hyperref}

\usepackage[accepted]{V2arXiv}

\usepackage{amsmath}
\usepackage{amssymb}
\usepackage{mathtools}
\usepackage{amsthm}

\usepackage[capitalize,noabbrev]{cleveref}

\theoremstyle{plain}

\theoremstyle{definition}

\theoremstyle{remark}

\usepackage[textsize=tiny]{todonotes}

\icmltitlerunning{Increased Compute Efficiency and the Diffusion of AI Capabilities}

\begin{document}
\onecolumn
\icmltitle{Increased Compute Efficiency and the Diffusion of AI Capabilities}



\icmlsetsymbol{equal}{*}

\begin{icmlauthorlist}
\icmlauthor{Konstantin Pilz}{equal,Georgetown}
\icmlauthor{Lennart Heim}{equal,GovAI}
\icmlauthor{Nicholas Brown}{RAND}
\end{icmlauthorlist}

\icmlaffiliation{GovAI}{Centre for the Governance of AI, Oxford, United Kingdom}
\icmlaffiliation{Georgetown}{Georgetown University, Washington D.C., United States}
\icmlaffiliation{RAND}{RAND, Santa Monica, CA, United States}

\icmlcorrespondingauthor{Lennart Heim}{lennart.heim@governance.ai}
\icmlcorrespondingauthor{Konstantin Pilz}{kfp15@georgetown.edu}


\vskip 0.3in

\makeatletter
\renewcommand{\ICML@appearing}{}
\renewcommand{\Notice@String}{}
\makeatother
\printAffiliationsAndNotice{\icmlEqualContribution} 

\begin{abstract}

Training advanced AI models requires large investments in computational resources, or \textit{compute}. 
Yet, as hardware innovation reduces the price of compute and algorithmic advances make its use more efficient, the cost of training an AI model to a given performance falls over time --- a concept we describe as \textit{increasing compute efficiency}.
We find that while an \textit{access effect} increases the number of actors who can train models to a given performance over time, a \textit{performance effect} simultaneously increases the performance available to each actor. 
This potentially enables large compute investors to pioneer new capabilities, maintaining a performance advantage even as capabilities diffuse.
Since large compute investors tend to develop new capabilities first, it will be particularly important that they share information about their AI models, evaluate them for emerging risks, and, more generally, make responsible development and release decisions.
Further, as compute efficiency increases, governments will need to prepare for a world where dangerous AI capabilities are widely available --- for instance, by developing defenses against harmful AI models or by actively intervening in the diffusion of particularly dangerous capabilities.

\end{abstract}

\clearpage
\section*{Executive Summary}\label{executive-summary}

The cost of training an AI model to a given level of performance falls
over time. In 2017, training a classifier to 93\% accuracy on ImageNet
cost over \$1,000; in 2021, it cost only \$5 — a reduction of over
99\%. What are the drivers of this trend and what does it imply for the
diffusion of AI capabilities?

Falling training costs stem from improvements in two key areas:

\begin{enumerate}
\item
  Advances in \textbf{hardware price performance} — as predicted by
  Moore's Law — increase the computational performance available per
  dollar, often measured in FLOP/s per \$. Between 2006 and
  2021, the price performance of AI accelerators doubled
  approximately every two years.

\item
  Advances in \textbf{algorithmic efficiency} decrease the number of computational operations needed to train an AI model to a given level of performance. For example, between 2012 and 2022, advances in
  image recognition algorithms halved the compute required for achieving
  93\% classification accuracy on the ImageNet dataset every nine
  months.
\end{enumerate}
\begin{figure}[H]
    \centering
    \includegraphics[width=0.8\linewidth]{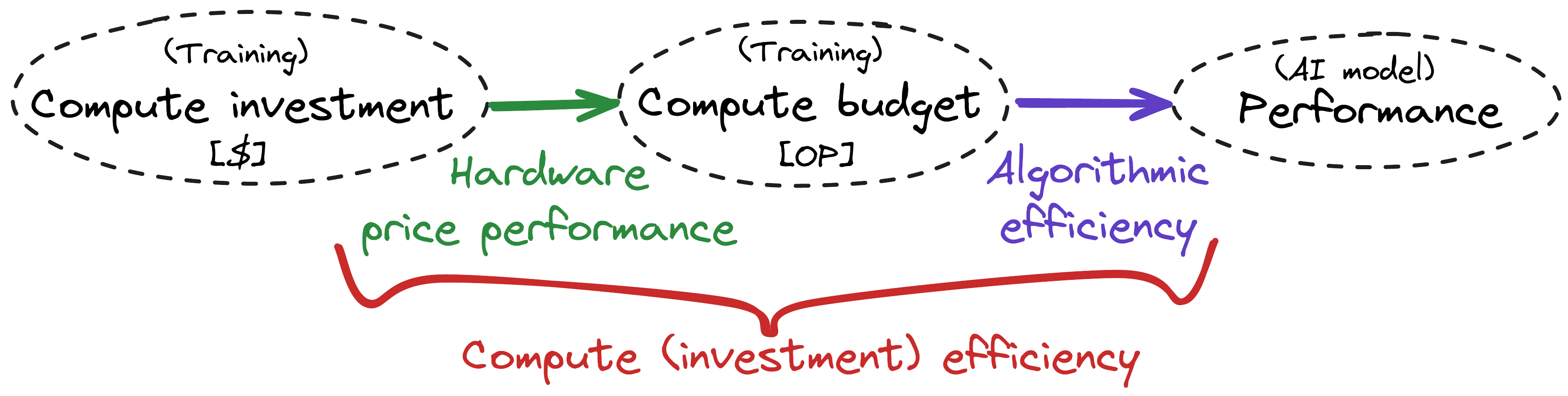}
    \caption{Compute efficiency relates the compute investment to the performance of an AI model.}
    \label{fig:5}
\end{figure}
To capture the combined impact of these factors, we introduce the concept of \textit{compute investment efficiency} —
abbreviated to \emph{\textit{compute efficiency}} — which describes how efficiently investments in training compute can be converted into AI model performance (\cref{fig:5}).\footnote{In the context of
  compute efficiency, \emph{AI model performance} refers to a directly
  measurable model property such as test loss for a language model or
  the cumulative reward for an RL agent.} Compute investment efficiency
determines the \emph{AI model performance} available with a given level of
investment in \emph{training compute}, provided sufficient training data (\cref{introduction}).

\textbf{Increasing compute efficiency has two main effects }(\cref{effects-of-compute-efficiency-increases})\textbf{:}
\protect\footnote{Our model and its implications are partly derived from \citet{tucker_social_2020}.}

\begin{figure}[!h]
    \centering
    \includegraphics[width=0.4\linewidth]{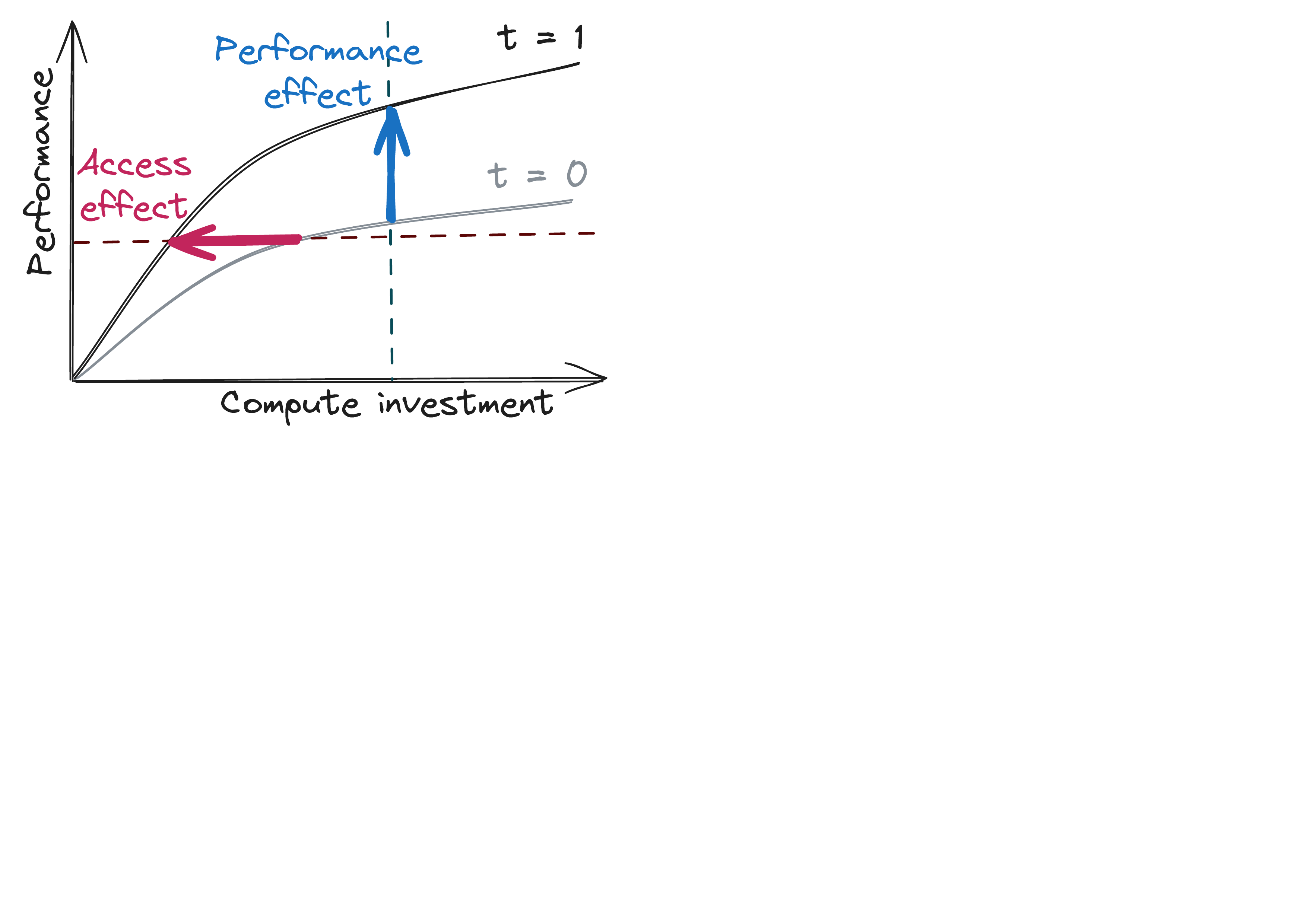}
    \caption{Compute efficiency improves between time $t = 0$ and $t = 1$, causing an access effect (red) and a performance effect (blue). \textit{Figures are merely conceptual and do not assert
  specific claims regarding the slopes of the curves.}}
    \label{fig:6}
\end{figure}

\begin{itemize}
\item \textbf{An access effect}: over time, access to a given level of
  performance requires less compute investment (\cref{fig:6}, red).
\item \textbf{A performance effect}: over time, a given level of compute
  investment enables increased performance (\cref{fig:6}, blue).
\end{itemize}

\textbf{If actors have equal access to compute efficiency improvements,
these effects have the following consequences} (\cref{consequences-of-the-effects-for-different-actors}):\protect
  
\begin{itemize}

  \item
  \textbf{Capabilities diffuse over time.} Due to the access effect, the
  investment required to reach a given performance level decreases over
  time, giving more actors the ability to reproduce capabilities
  previously restricted to large compute investors.
  \item
  \textbf{Large compute investors maintain a performance advantage even
  as capabilities diffuse.} Compute efficiency improvements that expand
  access to smaller actors simultaneously increase the performance
  available to large compute investors. Absent a ceiling on absolute
  performance, those actors maintain a performance advantage and thus continually discover novel capabilities first.

\end{itemize}

\textbf{We discuss some additional factors that influence the consequences of increasing compute efficiency} (\cref{implications-for-competitive-advantage}):
\begin{itemize}
    \item
  \textbf{Performance ceilings dampen the performance effect, reducing
  leaders' advantage.} Many AI applications have a ceiling on technical
  performance or real-world usefulness. For instance, handwritten digit
  classifiers have achieved above 99\% accuracy since the early 2000s,
  so further progress is insignificant. As leaders approach the ceiling,
  performance only marginally increases with improved compute
  efficiency, allowing smaller actors to catch up.
  \item
  \textbf{Winner-takes-all effects may allow leaders to entrench their
  lead even as they lose their performance advantage}. By initially
  developing the best-performing models, large compute investors may
  accrue a number of advantages unrelated to performance, such as
  network effects and economies of scale.

\item
  \textbf{In zero-sum competition, a relative performance advantage may
  grant outsized benefits.} If AI models directly compete, the developer
  of the leading model may reap disproportionate benefits even if their
  absolute performance advantage is small. Such disproportionate rewards
  may apply to AI models used
  in trading, law, or entertainment.
\end{itemize}

\textbf{Implications for Dangerous Capabilities} (\cref{implications-for-dangerous-capabilities})

\begin{itemize}
\item
  \textbf{Dangerous capabilities first appear in models trained by large compute investors} Future AI models could show dangerous capabilities, such as
  exploiting cybersecurity vulnerabilities, bioweapon design, or evading
  human control. Since these capabilities likely require high levels of
  performance, large compute investors most likely encounter them
  first.

\item
  \textbf{Dangerous capabilities proliferate over time.} As efficiency
  improvements increase the performance available with a given compute
  investment, more actors become able to train and develop models with
  potentially dangerous capabilities, complicating coordination and
  therefore increasing the chance of accidents and misuse.

\item
  \textbf{Deploying leading AI models defensively could strengthen
  resilience against misuse and accidents.} To counteract harm caused by weaker models, large compute investors may be able to use their more advanced models to create defensive tools. For example, cybersecurity tools
  powered by advanced models could find vulnerabilities before weaker
  models can exploit them. However, some domains, such as biotechnology,
  may greatly favor the offense, making it difficult to defend against
  dangerous deployments even with superior models.
  
\end{itemize}
\pagebreak
\begin{figure}[!ht] 
    \centering
    \includegraphics[width=0.7\linewidth]{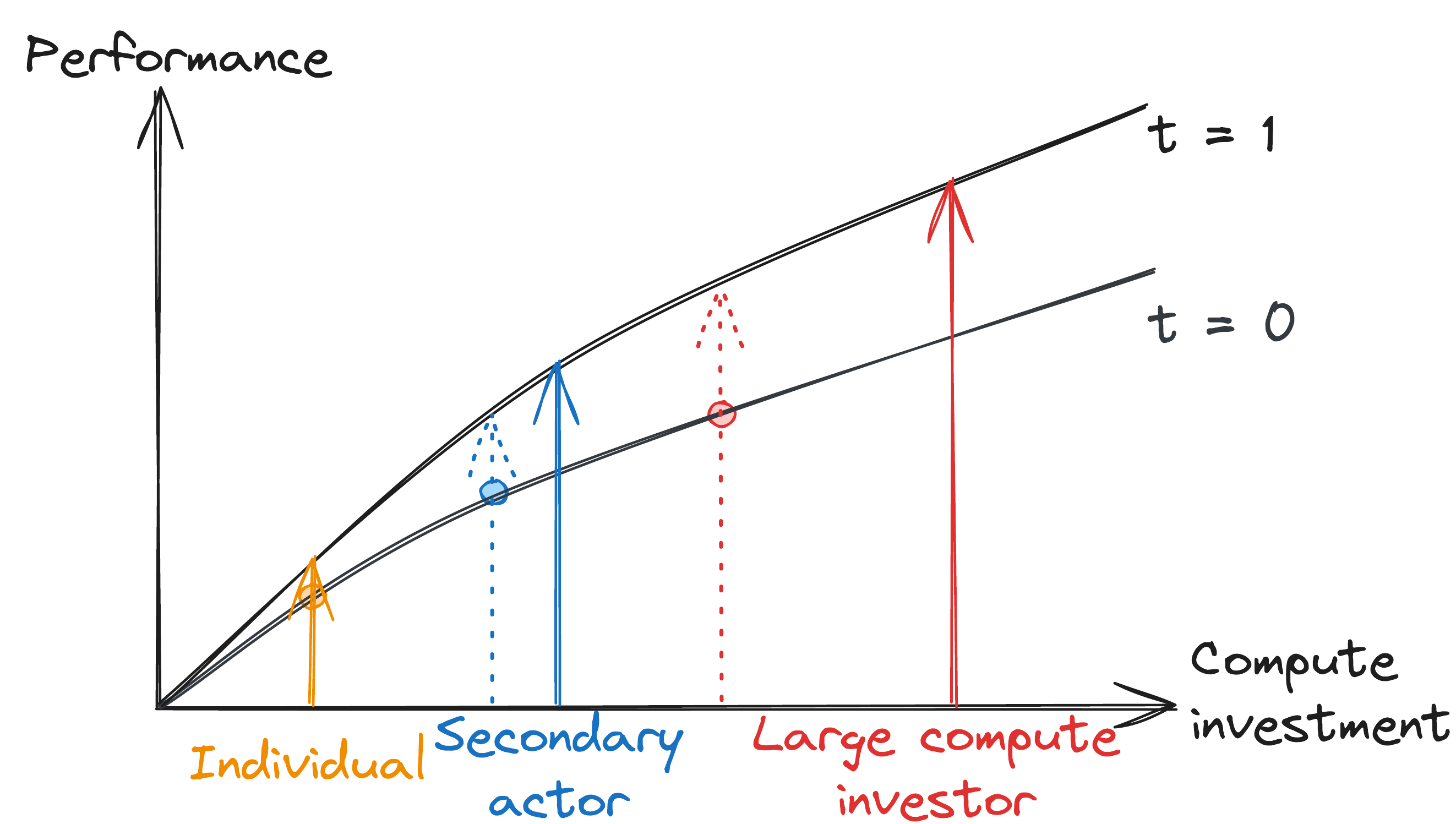}
    \caption{Compute investment scaling increases the performance lead of
large compute investors over time. The dashed arrows represent
performance attainable without investment scaling.}
    \label{fig:8}
\end{figure}

\textbf{Adjusting our model assumption to the real world further emphasizes the
outsized role of large compute investors} (\cref{additional-factors-influencing-the-distribution-of-capabilities}):

\begin{itemize}
\item
  \textbf{Leaders increase their performance advantage through
  increasing investments and proprietary algorithmic advancements.}
  Large compute investors have historically increased their compute
  investment significantly faster than others, widening the investment
  gap to smaller actors. Further, 
since they often employ a high number of talented researchers, large compute investors may develop proprietary hardware and algorithmic advances. These advances further widen the gap between large compute investors and other actors.
\item
  \textbf{Leaders can release their model parameters and, therefore,
  allow others to overcome compute investment barriers.} While product
  integrations and structured access protocols allow for continuous
  moderation, releasing model parameters causes irreversible
  proliferation, foregoing any control.
\end{itemize}
\textbf{Key Conclusions} (\cref{discussion}):
\begin{itemize}
\item
  \textbf{Oversight of large compute investors helps to address the most
  severe risks --- at least for a time.} If the most severe risks from AI development originate
  from the most capable models and their large-scale deployment, then
  regulating large-scale compute users can address the most severe
  risks. For instance, governments could mandate capability evaluation
  and risk assessment procedures for training large models, monitor
  large-scale deployments, and quickly intervene when models cause harm.
\item
  \textbf{Large compute investors should help mitigate risks from the
  proliferation of advanced capabilities.} The effectiveness of societal
  measures to mitigate harm from proliferation hinges on the time that
  passes between large compute investors discovering harmful capabilities
  and their proliferation to malicious or irresponsible actors. To
  effectively use this critical period, governments can implement
  information-sharing frameworks with large compute investors and require them to
  thoroughly evaluate the risks posed by capability proliferation.
\item
  \textbf{Governments should respond early to offense-dominant
  capabilities.} In the future, AI models of a given performance could
  develop heavily offense-dominant capabilities or become inherently
  uncontrollable. Governments should closely monitor the emergence of
  such capabilities and preemptively develop mechanisms that could
  restrict their proliferation.
\end{itemize}

\clearpage
\setcounter{tocdepth}{2}
\tableofcontents
\newpage
\section{Introduction}\label{introduction}

Over the past decade, computational resources (compute) emerged as a
major driver of rapid advances in the field of artificial intelligence
(AI). The amount of computational operations used in training the
largest AI models has doubled approximately every six months since 2010 \citep{sevilla_compute_2022}, enabling powerful new applications like high-quality image
generators, coding assistants, and conversational chatbots
\citep{yu_scaling_2022,chen_evaluating_2021,openai_gpt-4_2023}.
Researchers have discovered empirical scaling laws that formalize the
relationship between increased training compute and improved model
performance across numerous domains.\footnote{Scaling laws have been
  found in a variety domains, such as natural language processing
  \citep{kaplan_scaling_2020,hoffmann_training_2022}, protein structure
  prediction, acoustic modeling, and recommendation models \citep{chen_xtrimopglm_2023,droppo_scaling_2021,ardalani_understanding_2022}.} 
These scaling laws and the current pace of training compute growth suggest that compute will continue to contribute to increased model performance for the foreseeable future \citep{sevilla_compute_2022,villalobos_scaling_2023}.\footnote{Our model assumes that compute is a main determinant
  of an AI model's capabilities. However, note that, in reality,
  training AI models on large amounts of compute is a sophisticated
  engineering challenge that requires talented researchers. Thus, simply
  having access to large amounts of compute does not guarantee a
  powerful AI model. We discuss additional factors and limitations in \cref{additional-factors-influencing-the-distribution-of-capabilities}.}

\subsection{Falling Training Cost}

As compute budgets have surged, the cost of training an AI model to a
given performance has fallen continuously. In 2017, training an image
classifier to 93\% classification accuracy on the ImageNet dataset cost
over \$1,000; by 2021, it cost only \$5 — a reduction of over 99\% \citep[p. 97]{zhang2022}. In 2020, OpenAI's GPT-3 cost at least \$4.6
million in cloud compute to train
\citep{li2020}; two years later, the company Mosaic claimed to achieve the same
performance for a tenth of the price
\citep{venigalla2022}.

The cost of training falls due to two primary factors: (a) improvements
in the price performance of AI hardware and (b) improvements in the
efficiency of AI algorithms.

\subsection{Hardware Price
Performance}\label{hardware-price-performance}

\textit{Hardware price performance} determines the quantity of computational
resources available for a given monetary investment. Innovations in hardware
manufacturing and design increase price performance over time. For
example, consumer prices for Intel's processors fell by
42\% annually from 2004 to 2013, while their performance improved by 29\%
each year; both trends increased the computational performance available
per dollar
\citep{byrne2017}. In the context of AI accelerators, \citet{hobbhahn_trends_2023}
found that GPUs used in machine learning doubled in price
performance approximately every two years. In addition, increased specialization, such as hardware-supported lower-precision number formats, further increased
price performance by up to an order of magnitude
\citep{hobbhahn_trends_2023}.

 These trends are driven in part by a continued increase in the
 transistor density of chips, popularly known as Moore's
 Law \citep{roser2023a}, and by specialized chip design
 \citep{hobbhahn_trends_2023}. Beyond nominal chip performance, improvements in other
 factors like power and cooling can also lower the cost of operation and
 therefore improve price performance
 \citep{katal_energy_2023}.

\subsection{Algorithmic
Efficiency}\label{algorithmic-efficiency}

Falling AI training costs stem not just from more compute being
available per dollar but also from ongoing improvements in \textit{algorithmic
efficiency}. Algorithmic efficiency determines the amount of computational
resources required to execute a given algorithm. In the context of AI,
algorithmic efficiency identifies the compute budget required to train or run inference on a model providing a given level of performance. In this paper, we will focus on the former, the efficiency of training.

Due to innovations across the AI stack, algorithmic efficiency has
improved dramatically over the past decade. For example,
\citet{hernandez_measuring_2020} found that over the period of 2012 to 2019, improvements in image
classification algorithms had led to a 97.7\% reduction in the compute
required to train a classifier to the performance of AlexNet. \citet{erdil_algorithmic_2023} extended the analysis to 2022 and found an even faster rate of algorithmic improvement in image classification, in which compute requirements halved every nine
months.
Although limited analysis exists, algorithmic improvements have also impacted other fields. 
For instance, in language modeling, less compute-intensive models increasingly replicate the performance of previous, larger models~\citep{aiMixtralExperts2023, hoffmann_training_2022, venigalla2022}.


\subsection{Compute Investment
Efficiency}\label{compute-investment-efficiency}

\begin{figure}[h]
    \centering
    \includegraphics[width=1\linewidth]{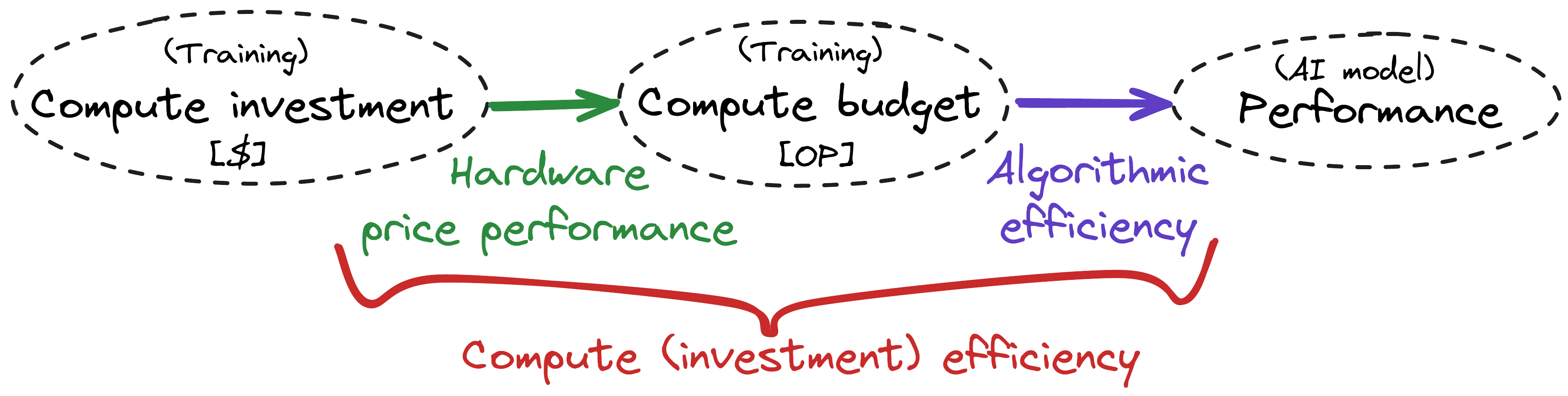}
    \caption{Hardware price performance is the conversion function between
the training compute investment in dollars and the training compute
budget in operations. Algorithmic efficiency is the subsequent
conversion function between the training compute budget and the
performance of the resulting AI model. \textit{Compute (investment) efficiency}
combines hardware price performance and algorithmic efficiency, relating
training compute investment to the performance of the resulting model.}
    \label{fig:1}
\end{figure}

We introduce the concept of \emph{compute investment efficiency} (abbreviated to \emph{compute efficiency}), to refer to the relationship
between the monetary spending on training compute and the performance of
the resulting AI model (\cref{fig:1}).\footnote{The model presented in this paper and many of its implications draw heavily from \cite{tucker_social_2020}
  ``Social and Governance Implications of Improved Data Efficiency''. We
  translate the concepts and implications discussed in the context of
  increased data efficiency to the realm of compute efficiency.}
Tracking compute efficiency over time captures advances in both the
hardware and software used for AI training, including those advancements
that depend on a mixture of hardware and software
breakthroughs.\footnote{For instance,
  \citet{hobbhahn_trends_2023} identify the recent shift from FP32 precision to
  tensor-FP16/8 as one of the primary drivers of AI hardware
  improvements.} For a formalization of our model, see \Cref{formal-model}.

In the following sections, we assess the impact of increasing compute efficiency on various actors and evaluate its implications for competitive advantage and
for the risks of dangerous capabilities. Finally, we provide a short
discussion of relevant insights for policymakers and anyone else seeking
to understand or respond to increased compute efficiency.

\section{Formal Model}\label{formal-model}

This section formalizes our model assumptions and defines key terms. 
In
the context of compute efficiency, the paper distinguishes an AI model's
performance from the resulting capabilities:

\begin{enumerate}
\item
  \emph{\textbf{Performance}} refers to a hard metric, such as the loss on a test dataset or the cumulative reward for an RL agent.
\item
  \emph{\textbf{Capabilities}} refer to a more qualitative metric, such
  as the problems an AI model can solve in the real world. They are,
  therefore, more difficult to quantify.
\end{enumerate}

For illustration, a language model that achieves a certain
\emph{performance} on next-word prediction may gain the
\emph{capability} to solve coding problems. While the performance of
models increases continuously, capabilities may emerge abruptly and are
more difficult to predict
\citep{ganguli_predictability_2022}.

Note that in our context, performance explicitly refers to \textit{direct
performance measures} rather than \textit{the model's performance on benchmarks}.
Depending on their nature, benchmarks can capture either the performance
of a model or its capabilities. For instance, the ImageNet benchmark
\citep{deng_imagenet_2009} directly evaluates the models' ability to
classify images, the task they were explicitly trained for. On the other
hand, diverse benchmarks such as Massive Multitask Language
Understanding (MMLU)
\citep{hendrycks_measuring_2021} primarily evaluate models' usefulness on real-world tasks such
as answering scientific exams and legal questions.

\subsection{Algorithmic Efficiency}

The performance $p$ achievable with a compute budget
$b$ depends on the algorithmic efficiency
$f_a$ (assuming sufficient access to
training data).
\begin{equation}
    p = f_a(b),
\end{equation}
where:
\begin{align*}
    p &= \text{Performance (accuracy of the AI model)}\\
    b &= \text{Compute budget (in operations)}\\
    f_a &= \text{Algorithmic efficiency}
\end{align*}
Algorithmic efficiency improves over time. Hence the same compute budget
yields improved performance:
\begin{equation}
    f_a^{t+1}(b)>f_a^{t}(b),
\end{equation}
where:
\begin{align*}
t &= \text{time}
\end{align*}

\subsection{Hardware Price Performance}

Meanwhile, the compute budget $b$ available for training
depends on the compute investment $i$ in dollars and the
hardware price performance $f_h$ that
converts the compute investment into the compute budget.
\begin{equation}
    b = f_h(i),
\end{equation}
where:
\begin{align*}
b &= \text{Compute budget (in operations)} \\
i &= \text{Compute investment (in $\$$)} \\
f_h &= \text{Hardware price performance (in
operations/$\$$)}    
\end{align*}
As hardware price performance improves over time, the same compute
investment yields a larger compute budget:
\begin{equation}
    f_h^{t+1}(i)>f_h^t(i)
\end{equation}
In other words, compute gets cheaper over time.

\subsection{Compute Investment Efficiency}

Compute investment efficiency $f_c$
combines hardware price performance $f_h$
and algorithmic efficiency $f_a$ into one
term. It thus determines the performance obtainable with a given
financial investment:
\begin{equation}
    f_c(i) = f_a\left(f_h(i)\right),
\end{equation}
where:
\begin{align*}
f_c &= \text{Compute investment efficiency}\\
i &= \text{Compute investment (in $\$$)}\\
f_h &= \text{Hardware price performance (in
operations / $\$$)}\\
f_a &= \text{Algorithmic efficiency}
\end{align*}

This simplifies the relationship between performance and investment to:
\begin{equation}
    p = f_c(i)
\end{equation}
where:
\begin{align*}
p &= \text{Performance} \\
f_c &= \text{Compute investment efficiency}\\
i &= \text{Compute investment (in $\$$)}
\end{align*}

Since both $f_a$ and
$f_h$ increase over time,
$f_c$ similarly increases over time.
\begin{equation}
    f_c^{t+1}(i) > f_c^t(i)
\end{equation}
In other words, the performance attainable with a given compute investment increases over time.

\section{Effects of Compute Efficiency Increases
}\label{effects-of-compute-efficiency-increases}

Similar to the effects of improved data efficiency in \citet{tucker_social_2020}, we model the impact of increased compute efficiency through two
effects: an \textit{access effect} and a \textit{performance effect}. 

\begin{figure}
    \centering
    \includegraphics[width=0.4\columnwidth]{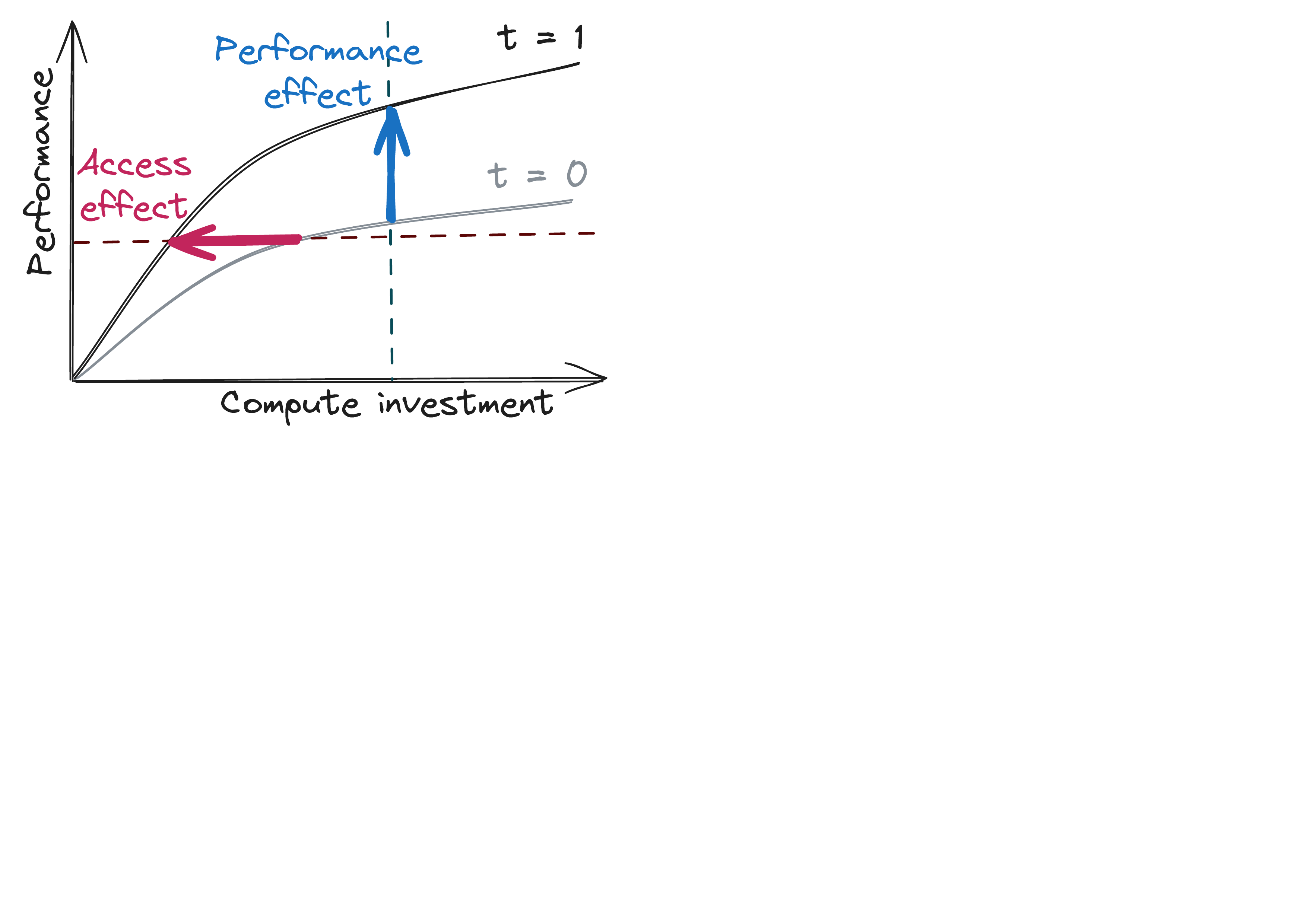}
    \caption{Compute efficiency improves between time $t = 0$ and $t = 1$, causing an access effect (red) and a performance effect (blue). \emph{Figures are conceptual and do not make empirical claims about the slope of the curve.}}
    \label{fig:2}
\end{figure}

\subsection{The Access Effect}\label{the-access-effect}

Over time, training a model to a given level of performance\footnote{We use \emph{performance} to refer to hard metrics such as test loss or the accuracy on a test dataset. We later use \emph{capabilities} to refer to qualitative metrics such as the types of problems an AI model can solve.} requires less compute investment. This \emph{access effect} corresponds to a leftward shift on the performance vs. investment graph (\cref{fig:2}, red), allowing increasingly many actors to access models with a given level of performance.

\subsection{The Performance
Effect}\label{the-performance-effect}

Simultaneously, a fixed level of compute investment allows actors to train models to a higher performance over time. This \emph{performance effect} corresponds to an upward shift in the performance vs. investment graph
(\cref{fig:2}, blue). Compute efficiency improvements thus grant any actor
access to increased performance.

\section{Consequences}\label{consequences}

In this section, we illustrate the 
effects of our model through an example with three types of actors. We
then assess what this implies for competitive
advantage and the discovery and proliferation of dangerous
capabilities.

\subsection{Consequences of the Effects for Different
Actors}\label{consequences-of-the-effects-for-different-actors}

To illustrate the impact of increased compute efficiency, we consider a
scenario with three groups of actors:
\footnote{
For simplicity, we make the following assumptions about these groups (each of which will be
relaxed in \cref{additional-factors-influencing-the-distribution-of-capabilities}):
(1) Actors maintain a constant level of compute investment over time,
(2) all actors have equal access to compute efficiency improvements, and
(3) actors only have access to the capabilities of the models they train themselves. See Appendix \ref{additional-factors-influencing-the-distribution-of-capabilities} for a discussion of these assumptions.
}
\begin{enumerate}
\item
  \textbf{Large compute investors}, who spend large amounts
  on training compute in order to advance the frontier of AI
  capabilities.
\item
  \textbf{Secondary actors,} who are able to make significant
  investments in training compute but who do \emph{not} attempt to
  advance the frontier. Examples include large companies that integrate
  some AI-based tools into their products and small companies developing
  AI models on a limited budget.
\item
  \textbf{Compute-limited actors}, such as individuals, small
  collaborative projects, or academic researchers who only have limited compute budgets. 
\end{enumerate}

We further assume that a model attains some relevant ``Capability $X$''
when it surpasses a given level of performance. We now
extrapolate repeated efficiency improvements and their effect on access
to ``Capability $X$''.

\begin{figure}
    \centering
    \includegraphics[width=0.7\linewidth]{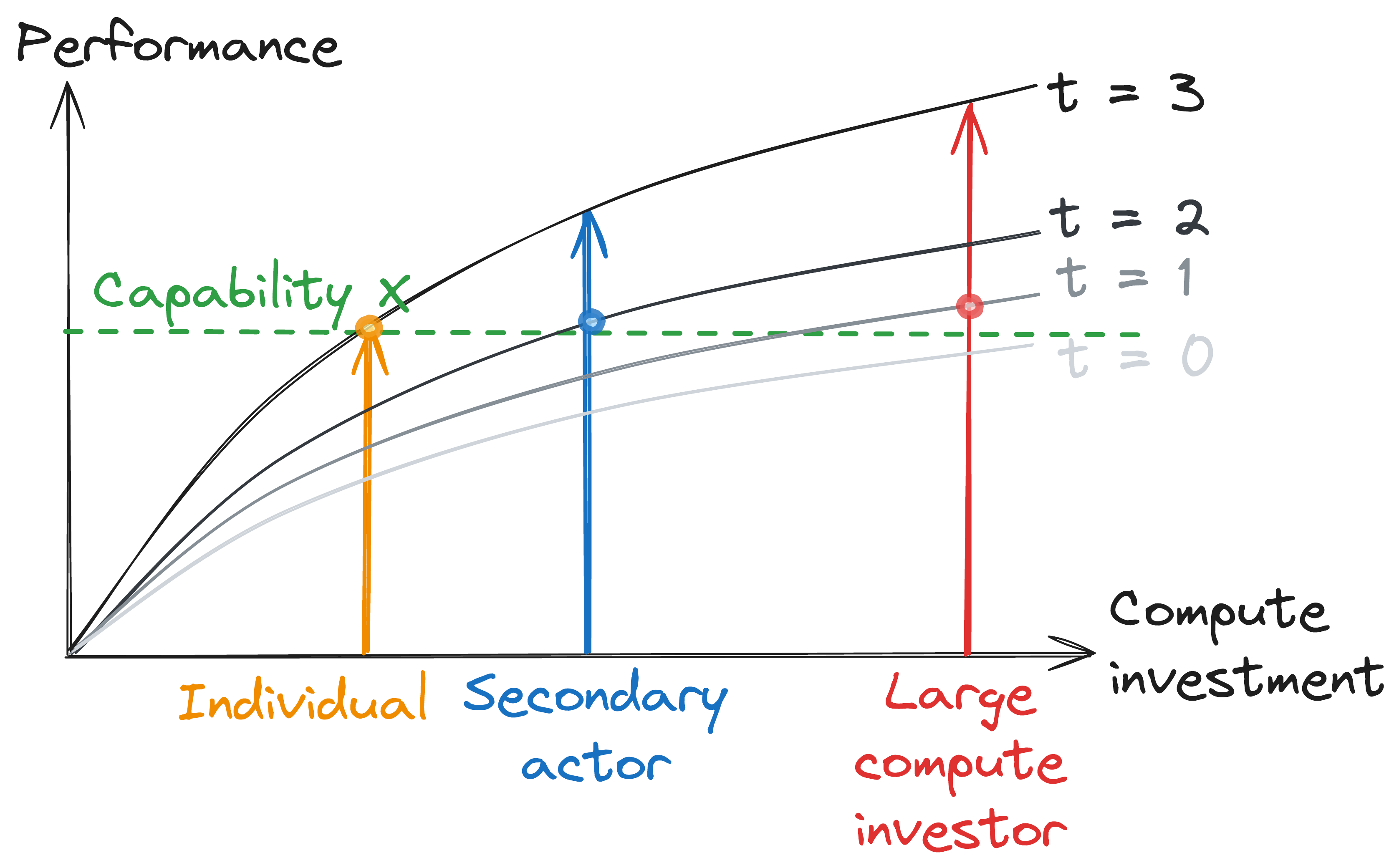}
    \caption{Diffusion of Capability $X$ over time among three illustrative
actors with varying levels of compute investment. 
 At time $t = 0$, no
 actor can train models with Capability $X$ with the performance enabled by their
 current level of compute investment. The large compute investor first
 achieves sufficient performance to attain Capability $X$ as compute
 efficiency increases at time $t = 1$. While performance also increases for
 the two smaller actors, it is still insufficient to cross the
 performance threshold. At time $t = 2$, further compute efficiency
 improvements also enable the secondary actor with its moderate compute
 budget to train a model with Capability $X$. Finally, after ongoing improvements at
 time $t = 3$, even the individual can afford the compute to train a model with
 Capability $X$. Meanwhile, increased compute efficiency already grants the
 secondary actor and the large compute investor access to a substantially higher
 level of performance.
    }
    \label{fig:3}
\end{figure}

This extrapolation suggests the following insights:

\textbf{Large compute investors are the first to discover novel
capabilities.} 

As the performance effect increases the performance
attainable to every actor, large compute investors are the first to
discover novel capabilities (\cref{fig:3}, $t = 1$). For example, OpenAI's
GPT-4 — the first training run to surpass $10^{25}$ FLOP \citep{epoch_ai_2023} —
showed previously unknown capabilities in mathematical reasoning,
coding, and theory of mind \citep{openai2023b,bubeck_sparks_2023}.

\textbf{Access to capabilities expands over time.} 

As compute
efficiency continues to improve, less and less investment is required to achieve a
given level of performance. Over time, this lets more and more
resource-limited actors recreate capabilities that were previously available only to large compute investors (\cref{fig:3}, $t = 2$).
Eventually, even actors with very limited compute investments, such as
research collectives or individuals, gain access to advanced capabilities (\cref{fig:3}, $t = 3$).

\textbf{Large compute investors may sustain their performance advantage
as access expands.} 

This is because the same efficiency increase that
lets a smaller investor access the previous state-of-the-art (SOTA)
often
increases
the current SOTA as well (\cref{fig:3}, $t = 3$).
Assuming it is possible to derive additional value from increased
performance, large investors will maintain an advantage. We explore this
topic in more detail in the following section.

\subsection{Implications for Competitive
Advantage}\label{implications-for-competitive-advantage}

Our model suggests several implications for the competitive advantage of
different actors in the AI market.

\subsubsection{Implications of Capability Diffusion}\label{capability-diffusion-enables-innovation}

With each efficiency increase, a broader range
of actors can train AI models to a given performance, resulting in a
wider variety of products. However, our model suggests that the impact on competition may often be limited because large compute investors benefit from the performance effect and retain an absolute advantage. Hence, smaller companies may face pressure to specialize, pursuing applications in areas with reduced competition from leaders.

For instance, various companies have recently developed smaller language
models that perform significantly below OpenAI's GPT-4 on general
benchmarks
\citep{noauthor_papers_2023} but outperform it in specialized niches, such as
finance \citep{wu_bloomberggpt_2023}, coding assistance
\citep{noauthor_ai_2023-2,chen_evaluating_2021}, or
role-playing for entertainment
\citep{noauthor_characterai_2023}.

When AI capabilities finally diffuse even to low-resource actors like small research collectives and individuals, the range of use cases explored may increase significantly. This is because
these groups have a broader range of interests than commercial entities
do, so this diffusion could represent a qualitative shift in AI applications \citep{besirogluComputeDivideMachine2024, ahmed_-democratization_2020}.

\subsubsection{Threshold Effects}\label{threshold-effects}

AI models sometimes demonstrate large improvements in capabilities even with only small increases in
compute investment. These \textit{performance thresholds} often
occur in large language models, including for tasks like three-digit
addition and general benchmarks like MMLU
\citep{ganguli_predictability_2022}.

Performance thresholds play a vital role in the usefulness of a model.
For example, language models trained with small amounts of compute
struggle to generate coherent sentences; with more compute, they
eventually cross the performance threshold of producing text of
sufficient quality to power chatbots and other applications.\footnote{Most performance thresholds are gradual. In the previous example, the capability to generate coherent text is not a discrete jump but rather a rapid but continuous improvement.}


Our model suggests that when a compute efficiency advance first unlocks a qualitatively new capability, large compute investors are likely to be the only actors able to leverage it. Until subsequent advances broaden access (see \cref{fig:3}), these actors will enjoy reduced competition or even a full monopoly. 

\phantomsection
\subsubsection{Performance Ceilings}\hypertarget{performance-ceilings}{}

As developers scale their training compute investment, they may
eventually encounter a region where further compute investment only
marginally (if at all) increases a model's usefulness for real-world
tasks. For example, the problem of identifying handwritten digits was
effectively solved in the early 2000s, with classifiers reaching over
99\% accuracy on the MNIST dataset
\citep{liu_handwritten_2003}. Similarly, face recognition technology has achieved impressively low error rates, down to 1:1000 on the VISA
dataset
\citep{noauthor_face_2023}.

\begin{figure}[!ht]
    \centering
    \includegraphics[width=0.6\linewidth]{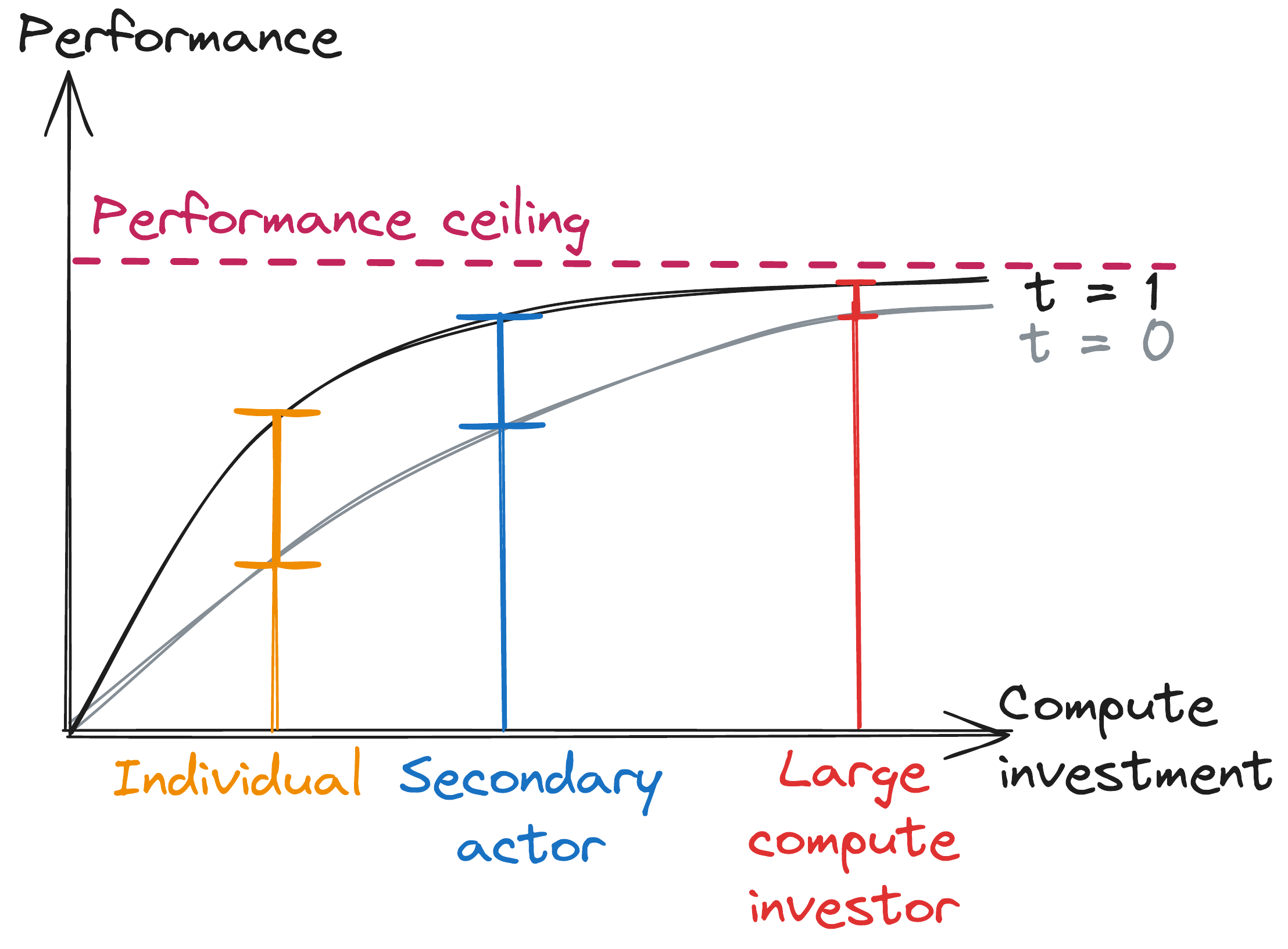}
    \caption{Performance ceilings dampen the performance effect for large compute investors, potentially allowing other actors to catch up.
    }
    \label{fig:4}
\end{figure}

 \subsubsection{Technical Performance Ceilings}

Performance ceilings can arise from either technical
limitations or real-world constraints. \citet{hestness_deep_2017} propose that, with sufficient training compute, an AI model may
encounter an irreducible error region where its accuracy can no longer
increase because the fixed training set and model architecture have
extracted all possible information relevant to the task. Yet, the authors find this error region only in a simplified toy experiment
and suggest that most AI applications are still far away from encountering it. 

 \subsubsection{Practical Performance Ceilings}

In practice, performance ceilings also occur when a model's performance continues to increase
but no longer contributes to advances in real-world capabilities. For
instance, in a voice-controlled application designed to recognize only
a fixed set of commands, improving the underlying model's accuracy on
less frequent words will not enhance its real-world functionality.
Similarly, a low-performance language model used in a content filter may
detect most harmful phrases already, so further advances improve its
usefulness only marginally.

 \subsubsection{Implications}

In our model, a performance ceiling creates an upper bound on achievable capability and thus dampens the performance effect, reducing the
benefits leaders get from improved compute efficiency. Since the access
effect remains unchanged, subsequent compute efficiency improvements let smaller
actors approach the performance of the leaders who have already reached the ceiling (\cref{fig:4}). The performance ceiling, therefore, distributes
capabilities evenly, and large compute investors slowly lose their
performance lead.

However, under some circumstances, leaders may entrench their initial
advantage and continue to dominate the field even as their performance
advantage diminishes. Such \emph{winner-take-all effects} occur, for instance, when
large compute investors integrate AI models into their established ecosystems and
thus create lock-in effects.\footnote{For instance, Google currently
  integrates its chatbot Bard into its search engine and the Google
  workspace
  \citep{noauthor_bard_2023}. Such integration may increase customers' switching costs, so
  they may continue using Google's AI model, even if competitors offer
  similar-performing models.} Alternatively, they may be able to use early profits to achieve economies of scale or apply exclusionary tactics to hamstring smaller competitors
\citep{kuchinke_exclusionary_2016}.

\subsection{Diminishing Returns}

Even without an apparent performance ceiling, AI models usually show
reducing improvements from increasing compute investment or compute
efficiency \citep{kaplan_scaling_2020}. If returns diminish quickly, the field may encounter an
effect similar to a performance ceiling, where the leaders' advantage
reduces with increasing compute efficiency.

\subsubsection{Zero-Sum Competition}\label{zero-sum-competition}

In certain domains, a relative performance advantage compared to
competitors can lead to outsized benefits, even if the absolute
performance difference is small.
This is most notably the case in settings with zero-sum competition,
\footnote{Note that while leaders in these markets capture outsized benefits, they may not constitute zero-sum games in an economic context. The overall market can still expand due to rising model capabilities, effectively resulting in a positive-sum game where increased competition drives market growth.} 
in which value is derived exclusively from performing better than an opponent. 
Zero-sum competition occurs in many domains, including algorithmic trading, in which a marginal speed advantage can let a trader exploit tiny market inefficiencies \citep{biais_equilibrium_2015}, law, in which the aim is to defeat an opponent's argument in court, and entertainment, in which competitors fight over a fixed pool of audience attention. Our model suggests that leaders may continue dominating in areas of zero-sum competition, even if their models only have a marginal performance advantage.

\subsection{Implications for Dangerous
Capabilities}\label{implications-for-dangerous-capabilities}

Today's AI models already cause harm, for instance, by
producing racist, sexist, or otherwise discriminating content or by
aiding the suppression of minorities or political views
\citep{mcgregor_preventing_2021,weidinger_ethical_2021,creemers_chinas_2018}. As models become more performant, they may develop capabilities
that pose increasingly acute risks to safety and security, such as by automating
cyberattacks
\citep{amodei2023}, aiding in the design of bioweapons
\citep{urbina_dual_2022,nelson2023}, generating targeted disinformation
\citep{sedova2021}, or irreversibly evading human control
\citep{carlsmith_is_2022,ngo_alignment_2023}.
Although some dangerous capabilities may first arise in highly
specialized models that require only small amounts of
compute, the majority likely first appear in compute-intensive general-purpose models \citep{anderljung_frontier_2023}.\footnote{Not all dangerous capabilities are concentrated in models requiring large amounts of compute. For instance, specialized models that predict host-pathogen interaction for viral proteins could significantly aid bioweapon development but may not require large amounts of training compute. For instance, AlphaFold, an advanced protein predictor, required only $10^{20}$ FLOP, three orders of magnitude less than OpenAI's GPT-3 developed in the same year \citep{senior_improved_2020,brown_language_2020,epochMachineLearningData2022}. Biological design tools similar to AlphaFold could play a crucial role in the development of bioweapons \citep{sandbrink_artificial_2023}.}

\subsubsection*{Discovery of Dangerous Capabilities}\label{large-compute-investors-encounter-dangerous-capabilities-first}

As discussed in \cref{implications-for-dangerous-capabilities}, we expect that large compute investors generally encounter novel capabilities first. This rule is likely to
hold equally for benign and dangerous capabilities. However, without rigorous testing,
developers may not immediately discover all dangerous capabilities of their models
\citep{shevlane_model_2023}. Instead, some may become apparent only after being broadly
available, including to potential malicious actors
\citep{anderljung_frontier_2023}.

\begin{figure}[ht]
    \centering
    \includegraphics[width=0.8\linewidth]{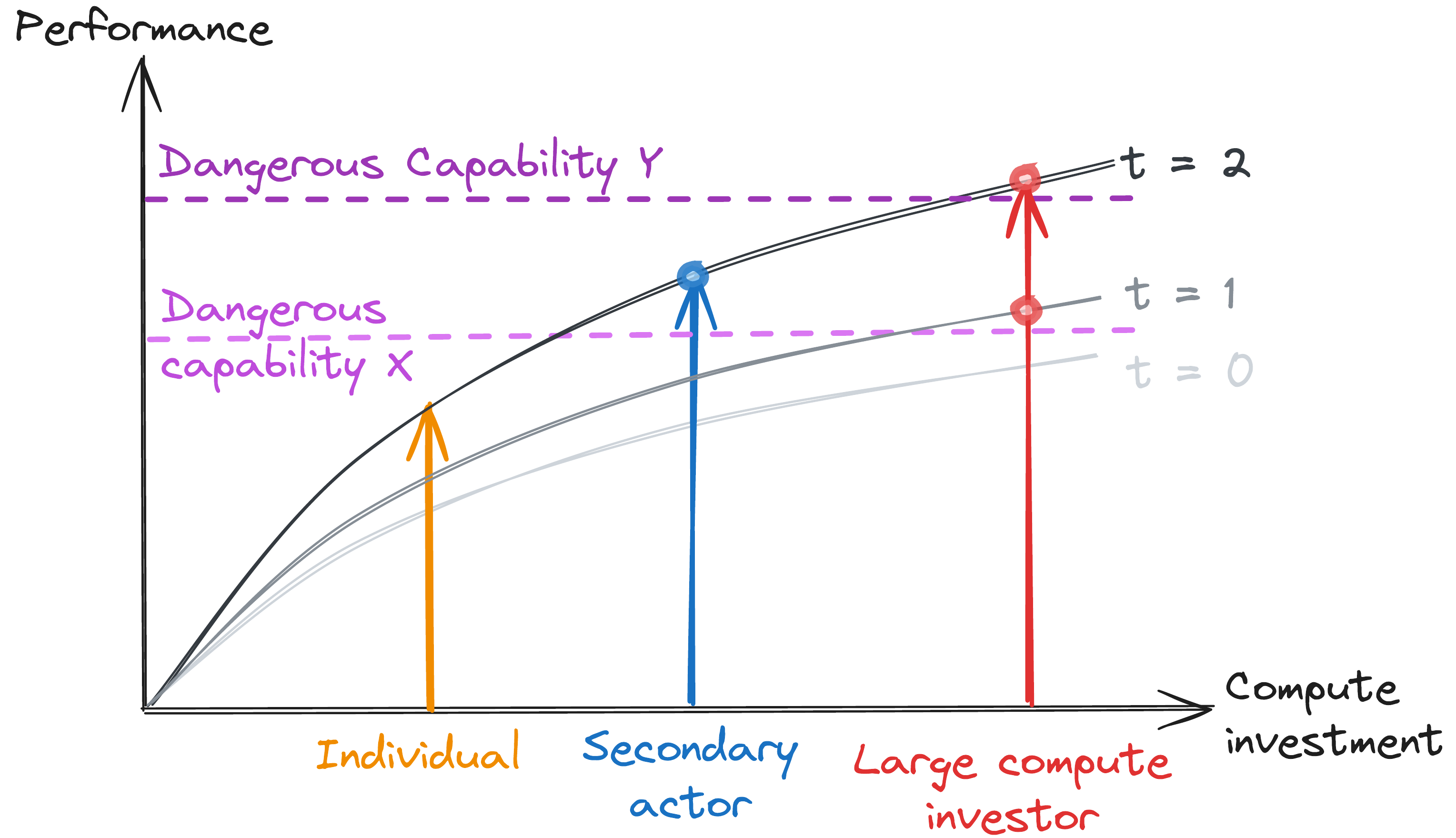}
    \caption{Example of three actors discovering novel dangerous capabilities $X$ and $Y$ at different points in time.
    }
    \label{fig:9}
\end{figure}


While only a small number of large compute investors will initially be able to train models that possess a novel dangerous capability X, compute efficiency improvements soon allow more and more actors to recreate such models (\Cref{fig:9}).
This complicates coordination and oversight, making it potentially
difficult to entirely avoid harmful deployment.

\subsubsection{Using Advanced Models for
Defense}\label{using-advanced-models-for-defense}

One way to mitigate risks from the proliferation of dangerous
capabilities is to invest in defensive measures. In particular, large compute investors may be able to use their
ongoing performance advantage to detect and address threats posed by
irresponsible or malicious actors
\citep[p. 25]{anderljung_frontier_2023}. For instance, large compute investors may offer
cybersecurity tools for automatic threat detection and response that
defend against attacks by less capable models developed by rogue actors
\citep{lohn2022}. Similarly, automated detection of disinformation
could limit the impact of AI on epistemic security
\citep{demartini2020}. Even as a model approaches a performance ceiling, large
compute investors may be able to provide effective defensive measures by
leveraging their superior quantity of inference compute (e.g., by
deploying more and/or faster model instances.\footnote{A model instance
  refers to a single copy of a particular AI model. Once a model is
  trained, the developer can run many such instances using the available
  supply of compute.})

\subsubsection{Importance of Offense-Defense
Balance}\label{importance-of-offense-defense-balance}

Nonetheless, some applications of AI capabilities may inherently favor
offensive use, making it difficult to defend against them even with
advantages in performance or scale of deployment.
\citet{garfinkel_how_2019} reviewed how the balance between offense and defense
scales with increased investment across a number of military scenarios
and found a high variance, with some technologies, such as missiles, likely favoring the offense. 
Some AI capabilities may similarly favor the offense. For instance, even a very capable protein predictor may not easily
find cures for toxic agents developed by a somewhat weaker model.

The offense-defense balance of advanced
capabilities may be particularly decisive when a field approaches a
performance ceiling. In this case, there may be little or no performance
advantage to count on; therefore, the feasibility of defending against
irresponsible or malicious use depends on whether the proliferating
capability fundamentally favors offense or defense.

Even without a ceiling, model proliferation in some particularly offense-dominant domains, such as biotechnology, may pose an unacceptable risk.
This is particularly concerning given the unpredictable nature of
AI progress makes it possible that future technologies could cause harm
on a large scale
\citep{bostrom_vulnerable_2019}.

\section{Additional Factors Influencing the Distribution of
Capabilities}\label{additional-factors-influencing-the-distribution-of-capabilities}

We discuss some limitations of our model and how adjusting it changes its implications.

\subsection{Compute Scaling}\label{compute-scaling}
\begin{figure}[!ht]
    \centering
    \includegraphics[width=0.8\linewidth]{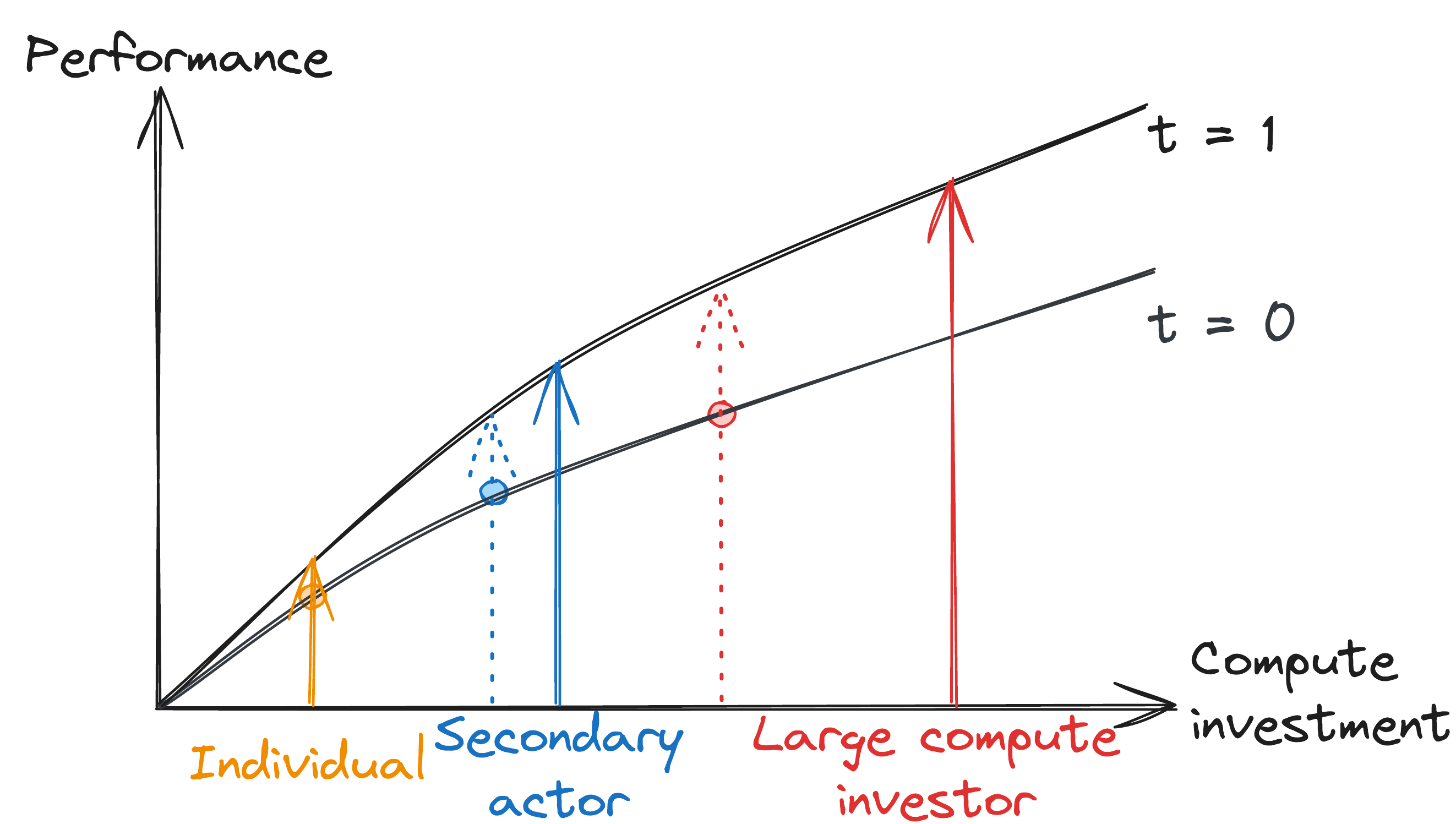}
    \caption{Illustration of the effect of compute scaling. The ability of
large compute investors to rapidly scale their investment greatly
increases the performance gap between them and other actors, who only
have limited capacity to scale. The dotted arrows represent the
performance each actor would achieve before scaling their investment.
The solid arrows represent the actual performance achieved due to
increased investment.
    }
    \label{fig:7}
\end{figure}

So far, our model has assumed that compute investment stays constant
over time. However, in reality, AI companies have massively increased
their investments, which rose thirtyfold between 2013 and 2021
\citep{roser_artificial_2023}. This exponential increase in investment is primarily
concentrated in large compute investors, allowing them to double their
training compute investment every two years
\citep{cottier_trends_2023}. For example, training GPT-3 cost OpenAI about \$4.6 million in
2020 \citep{li2020}, while GPT-4 reportedly cost up to \$100 million in 2023
\citep{knight_openais_2023}.

Although scaling compute presents a significant engineering
challenge,\footnote{For instance, parallelizing AI training across
  larger and larger clusters makes it hard to avoid bottlenecks that
  reduce utilization.} success grants large compute investors an
additional performance advantage and thus increases their lead over less
well-resourced actors (\cref{fig:7}). \citet{besirogluComputeDivideMachine2024}
identify the widening gap between the compute investment of industry and
academia as a compute divide
\citep{noauthor_computation_2023}. Although large compute investors can only
maintain exponential scaling for a limited time \citep{lohn2022a}, increasing investment will continue to extend their
lead for at least the near future.

\subsection{Data as a Bottleneck}

The increasing training compute budget stems from training models on
more data. However, there is a potential bottleneck if one were to run
out of data. While the availability of high-quality data is a crucial
determinant for performance in many specialized domains
\citep{budach_effects_2022}, it does
not currently limit scaling training compute for general-purpose models
that rely on non-specialized data
\citep{patel2023}. Yet, if current data scaling continues with these
increased compute training budgets, it could exhaust high-quality
language data by 2026
\citep{villalobos_will_2022}. A concentration of high-quality data within a specific actor
could considerably shift the competitive advantage, reducing the effects
of compute efficiency. However, algorithmic advances increase not only
compute efficiency, but also data efficiency --- the amount of data
required to achieve a given level of performance \citep{tucker_social_2020}.

\subsection{Unequal Distribution of Compute Efficiency
Improvements}\label{unequal-distribution-of-compute-efficiency-improvements}

Our model has assumed that all actors benefit from a compute efficiency
improvement. In reality, compute efficiency improvements are often
distributed unequally. One source of inequality is access to innovation
driven by the workforce. Large compute investors can afford large teams
and attract top researchers, often allowing them to discover algorithmic
advances first. Given the importance of such advances to their
profitability, frontier AI developers are increasingly reluctant to
publish those advances, given their relevance to competitive advantage
and their potential misuse. For example, in 2019, OpenAI published
details on the architecture, data, and algorithms used for GPT-2,
whereas in 2023, GPT-4's publication included almost no such information
\citep{radford2019,openai_gpt-4_2023}. As a result, algorithmic advances are increasingly concentrated
at frontier AI developers, diffusing only through leaks or parallel
innovation.\footnote{Over the last year, we
  observe that publication norms have entered a new phase. Frontier AI
  developers are reluctant to share even basic details of their models,
  such as architecture and compute used, making it challenging to
  estimate their internal compute efficiency. This further reduces the
  diffusion of compute efficiency improvements, which now primarily
  spread through employees switching companies.}\\
\strut \\
Additionally, several frontier AI companies or partners --- such as
Google, Meta, and Amazon
\citep{jouppi_tpu_2023,janardhan2023,noauthor_ai_2023-1} --- develop their own hardware. While some of these companies
offer external access to their custom hardware through cloud services,
internal teams likely incur lower costs and benefit from early access.
Custom hardware, therefore, potentially confers a unique price
performance advantage, further increasing the performance gap between
frontier AI labs and smaller actors.\footnote{Potentially, different
  frontier AI developers will design their own architectures, including
  optimized hardware, making it increasingly difficult to assess and
  compare their models.} \footnote{Should AI models become sufficiently
  capable to significantly enhance R\&D in hardware or algorithms, the
  performance advantage of large compute investors could allow them to
  deploy more and more capable automated workers that might further
  enhance the compute efficiency available to them.}

\subsection{Alternative Means of Diffusion: APIs and Release of
Model
Parameters}\label{alternative-means-of-diffusion-apis-and-release-of-model-parameters}

Our model has assumed that actors can only access capabilities of models
they can train themselves. However, this perspective misses other
avenues through which compute-limited actors can gain access to
SOTA AI models.

\subsection{API and Product Integrations}

Application programming interfaces (APIs) or AI models integrated into
products offer compute-limited actors access to otherwise inaccessible
advanced capabilities. Typically, users can only query such models
without being able to make changes to the model, but some APIs also
offer fine-tuning functionality, thus allowing users to partly retrain
custom models
\citep{peng2023}. In this access pattern, the model developer is able to
maintain control over the interaction, such as by charging for use and
implementing safeguards to prevent misuse and accidents.

\subsection{Public Release of Model Parameters}

Some major developers with significant compute investment have publicly
released the parameters of their advanced models. For example, Meta
released the parameters of Llama 2 for both research and commercial use
in July 2023 \citep{meta2023}.
This access pattern grants third parties unrestricted access to the
model, foregoing the ability to implement user guidelines or charge fees
for usage later. While this allows independent researchers to scrutinize
the models produced by large compute investors, it also significantly
increases the risk of misuse by irresponsible actors.
\citet{seger2023} suggest that structured access protocols may allow
independent scrutiny while limiting the risks from model parameters
release.

\subsection{Industrial Espionage}

Sophisticated actors like competitors, nation-states, or major terrorist
groups may try to infiltrate large compute investors and steal the
parameters of highly capable models. This poses risks not only to the
developer's business prospects but also to society if stolen models are
used for malicious purposes
\citep{nevo_securing_2023}.

\section{Discussion}\label{discussion}

We now discuss some of the broader implications of increasing compute efficiency. Given their importance, we focus on the implications for mitigating risks from the emergence and proliferation of dangerous capabilities. However, we also recognize the need for a better understanding of the implications of increasing compute efficiency on economics and competition and encourage further research on this topic.

Since large compute investors discover dangerous capabilities first, we argue their decisions deserve particular scrutiny. Yet, society also needs to prepare for the risks coming from the proliferation of these dangerous capabilities. We motivate research aiming at assessing and forecasting AI models' ability to cause harm and encourage defensive measures that could limit the negative consequences of proliferation.

\subsection{Responsibility of Large Compute Investors}\label{since-dangerous-capabilities-first-originate-from-large-compute-investors-they-should-face-particular-scrutiny}

Many of the most dangerous capabilities, such as hacking or social
manipulation, likely first arise in the highest-performing
general-purpose AI models, developed by a small number of large compute investors. 
In choosing to train such advanced models, large compute investors
have a responsibility to identify novel capabilities and the
risks they pose. To reliably detect potentially dangerous capabilities
of their models, developers can design and apply evaluations that test
whether models are capable of causing harm
\citep{liang_holistic_2023,shevlane_model_2023,kinniment2023}. Furthermore, developers can employ risk assessment
procedures that review more complex risks that may not directly arise
from a model's capabilities
\citep{novelli_taking_2023,koessler_risk_2023}.

While some frontier AI developers actively try to develop practices for
preventing harm caused by their AI models
\citep{anthropic2023}, incentives for industry-directed initiatives may not always
align with addressing the most severe risks. Therefore, governments
should scrutinize the decisions and methodologies AI
developers use, particularly targeting well-resourced labs that
concentrate on creating highly capable foundation models
\citep{anderljung_frontier_2023}.

\subsubsection{Overseeing Large-Scale Compute Infrastructure}

Assuming that the worst risks stem from the most capable AI models and
that these models' performance directly results from the size of their
training compute budget, then overseeing access to large-scale AI
compute clusters may present an effective method of regulation. Moreover, the amount of
compute available determines not only the trained model's capabilities
but also the number of model instances the developer can deploy.
Controlling access to large-scale compute resources could thus allow
governments not only to monitor the most advanced capabilities but also
to quickly address harms caused by the large-scale deployment of
potentially dangerous models
\citep{obrien2023a, egan_oversight_2023}.

 Currently, a large fraction of all AI compute likely belongs to a small
 number of cloud providers that offer their resources as a service
 \citep{richter2023}. These companies have already occasionally restricted harmful
 services, such as AWS banning the far-right social media site Parler
 from using its infrastructure after the January 6 attack on the US
 Capitol
 \citep{fitzpatrick2021}. To address proliferation risks, large cloud providers could
 introduce more extensive Know Your Customer (KYC) regimes, monitor
 suspicious activity, and develop methods to quickly shut down AI models
 that currently cause harm
 \citep{egan_oversight_2023,obrien2023a}. Besides avoiding harm, deployment oversight could also become a
 tool for attributing harms caused and holding actors liable
 \citep{anderljung_protecting_2023,buiten_law_2023}.

\subsection{Addressing Proliferation}\label{as-capabilities-proliferate-widely-addressing-the-harms-they-cause-becomes-increasingly-challenging}

As the access effect gives more and more actors the ability to reproduce models with dangerous capabilities using smaller and smaller amounts of compute it
becomes increasingly challenging to regulate the creation of such models.

AI researchers can help prepare society for this proliferation of dangerous capabilities by continually assessing SOTA AI models for their misuse potential and developing benchmarks and tests that could aid governments in determining the current and future risks associated with various AI models. 

Frontier AI developers may play a particularly crucial role in identifying and mitigating
risks associated with the proliferation of advanced model capabilities.
Given their knowledge advantage from developing the most capable models
and their research capacity due to designated safety and ethics teams,
these labs are in a strong position to investigate proliferation risks.
To allow society to prepare for the impact of proliferation, governments
may mandate frontier AI developers to share information about their
advanced models publicly or with selected government organizations
\citep{mulani2023,departmentforscienceinnovationandtechnology2023}. With insights into possible risks, governments could
then evaluate various measures aimed at either preventing model
proliferation or addressing its risks effectively.

Besides limiting the harmful use of proliferated models, governments may
also increase oversight of crucial inputs that allow malicious actors to
cause harm based on knowledge provided by these models
\citep{anderljung_protecting_2023}. For instance, controls on laboratory equipment and DNA
synthesis screening could prevent malicious actors from creating
dangerous pathogens even if the knowledge required is widely available
\citep{house_executive_2023,dieuliis_options_2017}.

\subsubsection{Using Advanced Models for
Defense}\label{using-advanced-models-for-defense-1}

Due to the performance effect, large compute investors continually create the most powerful models. The resulting performance advantage potentially
enables them to deploy AI models defensively to counteract risks caused by
proliferation. However, the feasibility of defending depends on several
crucial factors, such as the offense-defense balance, the gap between
leaders' and proliferated models, and the regulatory environment.

\subsubsection{Offense-Defense Balance}

As discussed in
\cref{implications-for-dangerous-capabilities}, the offense-defense balance may vary greatly between different
areas. Should proliferating AI capabilities significantly favor the
offense, it may become infeasible to use other models for defense, even
if the latter are more advanced and deployed on a larger scale. To
assess the feasibility of defensive solutions ahead of time, regulators
may need to continuously monitor current capabilities and conduct
experiments. However, assessing and experimenting with advanced models
will likely require frontier AI developers to securely share their
advanced models with trusted researchers, potentially enabled by secure
structured access protocols
\citep{seger2023,bucknall2023}.

\subsubsection{Proportion of Leaders' Performance Advantage}

Besides the offense-defense balance, the effectiveness of using advanced
AI models for defense also depends on the performance gap between the
defensive models and the proliferated capabilities.

The gap is largest when
defending against compute-limited actors like individuals, but
well-resourced actors like authoritarian states or irresponsible
companies could train AI models much closer in performance to the
frontier. As capabilities approach performance ceilings or diminishing
returns set in, the gap shrinks, limiting defense feasibility. However,
even near a ceiling, leaders can potentially leverage their large
compute investment and access to compute in order to deploy a large
number of model instances for defense and thus maintain a substantial
advantage over less well-resourced actors using proliferated models.

\subsubsection{Regulatory Environment}

Beyond just technical feasibility, the effectiveness of defensive AI
strategies may also depend on regulatory frameworks. Large compute
investors may not develop defensive measures by themselves, so
governments may need to contract them or initiate collaborative
projects. Moreover, certain defense approaches may require explicit
regulatory permission in order to be active in sensitive domains. For
example, cybersecurity defenses might need direct access to sensitive
networks and permission to make real-time decisions autonomously without
human oversight — which brings new risks. Similarly, protecting
individuals from social manipulation through AI may require monitoring
private communications, raising significant privacy concerns.
Governments, therefore, may need to develop and enforce standards that
preserve privacy in these defensive models while still giving them
sufficient permissions to ensure their effectiveness.

\subsection{Limiting Development and Proliferation of Dangerous
Capabilities}\label{limiting-development-and-diffusion-of-dangerous-capabilities}
If AI models strongly favor offensive applications or are an inherent
threat due to uncontrollability, or if defensive solutions are too
invasive, governments may have to limit the proliferation of such
capabilities in the first place. Given the difficulty in accurately
predicting these factors beforehand, it would be prudent to establish
precautionary mechanisms to manage potential risks.

\subsubsection{Misaligned AI Models}

So far, we primarily discussed how to address the misuse of advanced AI
models. However, an increasing number of AI researchers emphasize the
possibility that future, highly capable general AI models could pursue
autonomous goals, thus posing an inherent threat
\citep{carlsmith_is_2022,ngo_alignment_2023,russell_human_2019,centerforaisafety2023}. They argue it will be difficult to ensure such models are safe
and controllable. Some
frontier AI labs have already established designated research teams
focused on developing solutions for the safety of autonomous AI models
with near or above human-level capabilities
\citep{leike2023,anthropic2023a}. 

First, if advanced AI models are difficult to control, it could be hard
to leverage them for defense. Additionally, the gap between frontier AI
developers and others could narrow if frontier AI developers prioritize
safety over capability advances. Even if frontier AI developers
accurately assess that a model is uncontrollable and avoid deploying it,
increasing compute efficiency would still diffuse the ability to create such models over time, greatly
increasing the risk that someone eventually creates an uncontrollable model. 

\subsubsection{Balancing Scrutiny and
Proliferation Risks}

The critical period between the first discovery of a dangerous capability and its
proliferation to malicious actors is crucial for developing societal
resilience through regulation or defensive solutions. Hence, developers
should be cautious about hastening the diffusion of advanced AI models.
In particular, publishing the parameters of advanced AI models causes the
immediate, irreversible diffusion of the model's capabilities.
Rather than allowing anyone unrestricted access to advanced models,
frontier AI developers should introduce structured access procedures to
provide their model parameters only to responsible researchers
\citep{seger2023,solaiman_gradient_2023,shevlane_structured_2022,bucknall2023}.

\subsubsection{Coordinated Pausing}

 Should AI capabilities reach an unacceptable level of danger or models
 become too difficult to align, developers may have to halt further
 progress until they can appropriately mitigate these risks.
 \citet{alaga2023} suggest that frontier AI developers could
 collectively pause the development of highly capable models when they
 encounter capabilities deemed too dangerous. This pause, however, is
 likely only effective when few actors have access to such capabilities.
 If an extended pause is needed, the access effect will gradually enable
 more actors to develop dangerous models. In this scenario, preventing a
 growing number of entities from pursuing such development may require
 global coordination and strict oversight of compute resources. Yet, even
 with these measures, fully controlling the spread and use of these
 technologies may remain an elusive goal.

\section{Conclusion}\label{conclusion}

We found that increased compute efficiency results both in an access
effect — making a given capability more widely available — and a performance
effect — enabling a higher performance level for a given compute
investment. The impacts of the two effects depend on how a model's performance
converts to usefulness in real-world tasks. Threshold effects advantage
large compute investors, who are the first to discover qualitatively new
capabilities. Meanwhile, performance ceilings reduce the gap between
leading and lagging actors, whereas, in zero-sum competition, even a
marginal advantage allows outsized benefits for the leaders.

Our model suggests that large compute investors are at the frontier of
capabilities and, therefore, likely the first to discover dangerous
capabilities. To adequately address these capabilities, large compute
investors should implement extensive capability evaluation and risk
assessment procedures.

As compute efficiency increases, dangerous capabilities proliferate to
an increasing number of actors. The effectiveness of societal measures
to mitigate harm from this proliferation hinges on the duration between
frontier AI labs disclosing these capabilities and their proliferation
to malicious or irresponsible actors. To extend this critical period,
governments should implement information-sharing frameworks with leading
AI labs and thoroughly assess the risks of proliferation. Such assessments may necessitate access to advanced models by governments or
independent bodies.

Once informed about dangerous capabilities, governments should begin
increasing societal resilience against them. This can involve
restricting certain inputs required to cause harm, such as laboratory
equipment or compounds needed to develop dangerous pathogens.
Furthermore, governments can contract or coordinate with leading AI
developers to use their advanced models in defensive solutions that
address risks caused by proliferation. However, the feasibility of AI
models used for defense critically depends on the performance difference
between the defending model and the proliferated one, as well as the
fundamental offense-defense balance in the field.

Although increasing compute efficiency makes AI capabilities more widely
available over time, regulating access to large-scale compute clusters
can still increase oversight. Specifically, monitoring the largest
compute clusters allows monitoring the most extensive training runs,
which likely produce the models with the most advanced and potentially dangerous AI
capabilities.

Researchers working to ensure AI has a positive impact on society could direct their efforts toward evaluating model capabilities to detect potentially dangerous capabilities as early as possible. They could also develop benchmarks that reliably track AI progress over time and thus allow governments to anticipate dangerous capabilities ahead of time. Furthermore, AI researchers could prioritize research in fields that increase the resilience against these AI capabilities and enable the defensive use of AI models, particularly if insufficient incentives exist for frontier developers to do such research. 

However, both compute oversight and defensive solutions may be
inadequate should sufficiently dangerous capabilities arise, such as AI
models enabling individuals to create significant harm or if AI
models of a given performance are inherently uncontrollable. Governments
may need to preemptively develop mechanisms that could restrict the
development and proliferation of intolerably dangerous models, such as moratoria on specific research
areas to avoid large-scale harm.

\clearpage
\section*{Acknowledgements}
We thank Aaron Tucker, Markus Anderljung, and Allan Dafoe for their paper
``Social and Governance Implications of Improved Data Efficiency'' \citep{tucker_social_2020}
which inspired this paper and first introduced many of the ideas
presented here. We would like to express our thanks to the people who
have offered feedback and input on the ideas in this paper, including
Markus Anderljung, Ben Garfinkel, Ben Harack, and many other GovAI team
members. We thank Wes Cowley for copy editing and José Medina for
formatting assistance.


\bibliography{references}

\begin{thebibliography}{97}
\providecommand{\natexlab}[1]{#1}
\providecommand{\url}[1]{\texttt{#1}}
\expandafter\ifx\csname urlstyle\endcsname\relax
  \providecommand{\doi}[1]{doi: #1}\else
  \providecommand{\doi}{doi: \begingroup \urlstyle{rm}\Url}\fi

\bibitem[Ahmed \& Wahed(2020)Ahmed and Wahed]{ahmed_-democratization_2020}
Ahmed, N. and Wahed, M.
\newblock The {De}-democratization of {AI}: {Deep} {Learning} and the {Compute} {Divide} in {Artificial} {Intelligence} {Research}, October 2020.
\newblock URL \url{http://arxiv.org/abs/2010.15581}.
\newblock Issue: arXiv:2010.15581 arXiv:2010.15581 [cs].

\bibitem[Alaga \& Schuett(2023)Alaga and Schuett]{alaga2023}
Alaga, J. and Schuett, J.
\newblock Coordinated {Pausing}: {An} {Evaluation}-{Based} {Coordination} {Scheme} for {Frontier} {AI} {Developers} {\textbar} {GovAI}.
\newblock Research {Paper}, Centre for the Governance of AI, November 2023.
\newblock URL \url{https://www.governance.ai/research-paper/coordinated-pausing-evaluation-based-scheme}.

\bibitem[Amodei(2023)]{amodei2023}
Amodei, D.
\newblock Written {Testimony} of {Dario} {Amodei}, {Ph}.{D}. {Co}-{Founder} and {CEO}, {Anthropic}, July 2023.
\newblock URL \url{https://www.judiciary.senate.gov/imo/media/doc/2023-07-26_-_testimony_-_amodei.pdf}.
\newblock Subcommittee on Privacy, Technology, and the Law, United States Senate.

\bibitem[Anderljung \& Hazell(2023)Anderljung and Hazell]{anderljung_protecting_2023}
Anderljung, M. and Hazell, J.
\newblock Protecting {Society} from {AI} {Misuse}: {When} are {Restrictions} on {Capabilities} {Warranted}?, March 2023.
\newblock URL \url{http://arxiv.org/abs/2303.09377}.
\newblock Issue: arXiv:2303.09377 arXiv:2303.09377 [cs].

\bibitem[Anderljung et~al.(2023)Anderljung, Barnhart, Korinek, Leung, O'Keefe, Whittlestone, Avin, Brundage, Bullock, Cass-Beggs, Chang, Collins, Fist, Hadfield, Hayes, Ho, Hooker, Horvitz, Kolt, Schuett, Shavit, Siddarth, Trager, and Wolf]{anderljung_frontier_2023}
Anderljung, M., Barnhart, J., Korinek, A., Leung, J., O'Keefe, C., Whittlestone, J., Avin, S., Brundage, M., Bullock, J., Cass-Beggs, D., Chang, B., Collins, T., Fist, T., Hadfield, G., Hayes, A., Ho, L., Hooker, S., Horvitz, E., Kolt, N., Schuett, J., Shavit, Y., Siddarth, D., Trager, R., and Wolf, K.
\newblock Frontier {AI} {Regulation}: {Managing} {Emerging} {Risks} to {Public} {Safety}, November 2023.
\newblock URL \url{http://arxiv.org/abs/2307.03718}.
\newblock Issue: arXiv:2307.03718 arXiv:2307.03718 [cs].

\bibitem[Anthropic(2023{\natexlab{a}})]{anthropic2023}
Anthropic.
\newblock Anthropic's {Responsible} {Scaling} {Policy}, September 2023{\natexlab{a}}.
\newblock URL \url{https://www.anthropic.com/index/anthropics-responsible-scaling-policy}.

\bibitem[Anthropic(2023{\natexlab{b}})]{anthropic2023a}
Anthropic.
\newblock Core {Views} on {AI} {Safety}: {When}, {Why}, {What}, and {How}, March 2023{\natexlab{b}}.
\newblock URL \url{https://www.anthropic.com/index/core-views-on-ai-safety}.

\bibitem[Ardalani et~al.(2022)Ardalani, Wu, Chen, Bhushanam, and Aziz]{ardalani_understanding_2022}
Ardalani, N., Wu, C.-J., Chen, Z., Bhushanam, B., and Aziz, A.
\newblock Understanding {Scaling} {Laws} for {Recommendation} {Models}, August 2022.
\newblock URL \url{http://arxiv.org/abs/2208.08489}.
\newblock Issue: arXiv:2208.08489 arXiv:2208.08489 [cs].

\bibitem[AWS(2023{\natexlab{a}})]{noauthor_ai_2023-1}
AWS.
\newblock Ai {Accelerator} - {AWS} {Trainium} - {AWS}, November 2023{\natexlab{a}}.
\newblock URL \url{https://aws.amazon.com/machine-learning/trainium/}.

\bibitem[AWS(2023{\natexlab{b}})]{noauthor_ai_2023-2}
AWS.
\newblock {AI} {Code} {Generator} – {Amazon} {CodeWhisperer} – {AWS}, November 2023{\natexlab{b}}.
\newblock URL \url{https://aws.amazon.com/codewhisperer/}.

\bibitem[Besiroglu et~al.(2024)Besiroglu, Bergerson, Michael, Heim, Luo, and Thompson]{besirogluComputeDivideMachine2024}
Besiroglu, T., Bergerson, S.~A., Michael, A., Heim, L., Luo, X., and Thompson, N.
\newblock The {{Compute Divide}} in {{Machine Learning}}: {{A Threat}} to {{Academic Contribution}} and {{Scrutiny}}?, January 2024.

\bibitem[Biais et~al.(2015)Biais, Foucault, and Moinas]{biais_equilibrium_2015}
Biais, B., Foucault, T., and Moinas, S.
\newblock Equilibrium fast trading.
\newblock \emph{Journal of Financial Economics}, 116\penalty0 (2):\penalty0 292--313, May 2015.
\newblock ISSN 0304-405X.
\newblock \doi{10.1016/j.jfineco.2015.03.004}.
\newblock URL \url{https://www.sciencedirect.com/science/article/pii/S0304405X15000288}.
\newblock Number: 2.

\bibitem[Bostrom(2019)]{bostrom_vulnerable_2019}
Bostrom, N.
\newblock The {Vulnerable} {World} {Hypothesis}.
\newblock \emph{Global Policy}, 10\penalty0 (4):\penalty0 455--476, November 2019.
\newblock ISSN 1758-5880, 1758-5899.
\newblock \doi{10.1111/1758-5899.12718}.
\newblock URL \url{https://onlinelibrary.wiley.com/doi/10.1111/1758-5899.12718}.
\newblock Number: 4.

\bibitem[Brown et~al.(2020)Brown, Mann, Ryder, Subbiah, Kaplan, Dhariwal, Neelakantan, Shyam, Sastry, Askell, Agarwal, Herbert-Voss, Krueger, Henighan, Child, Ramesh, Ziegler, Wu, Winter, Hesse, Chen, Sigler, Litwin, Gray, Chess, Clark, Berner, McCandlish, Radford, Sutskever, and Amodei]{brown_language_2020}
Brown, T.~B., Mann, B., Ryder, N., Subbiah, M., Kaplan, J., Dhariwal, P., Neelakantan, A., Shyam, P., Sastry, G., Askell, A., Agarwal, S., Herbert-Voss, A., Krueger, G., Henighan, T., Child, R., Ramesh, A., Ziegler, D.~M., Wu, J., Winter, C., Hesse, C., Chen, M., Sigler, E., Litwin, M., Gray, S., Chess, B., Clark, J., Berner, C., McCandlish, S., Radford, A., Sutskever, I., and Amodei, D.
\newblock Language {Models} are {Few}-{Shot} {Learners}, July 2020.
\newblock URL \url{http://arxiv.org/abs/2005.14165}.
\newblock Issue: arXiv:2005.14165 arXiv:2005.14165 [cs].

\bibitem[Bubeck et~al.(2023)Bubeck, Chandrasekaran, Eldan, Gehrke, Horvitz, Kamar, Lee, Lee, Li, Lundberg, Nori, Palangi, Ribeiro, and Zhang]{bubeck_sparks_2023}
Bubeck, S., Chandrasekaran, V., Eldan, R., Gehrke, J., Horvitz, E., Kamar, E., Lee, P., Lee, Y.~T., Li, Y., Lundberg, S., Nori, H., Palangi, H., Ribeiro, M.~T., and Zhang, Y.
\newblock Sparks of {Artificial} {General} {Intelligence}: {Early} experiments with {GPT}-4, April 2023.
\newblock URL \url{http://arxiv.org/abs/2303.12712}.
\newblock Issue: arXiv:2303.12712 arXiv:2303.12712 [cs].

\bibitem[Bucknall \& Trager(2023)Bucknall and Trager]{bucknall2023}
Bucknall, B.~S. and Trager, R.~F.
\newblock Structured access for third-party research on frontier {AI} models: {Investigating} researchers’ model access requirements.
\newblock Whitepaper, Oxford Martin AI Governance Initiative, Centre for the Governance of AI, October 2023.
\newblock URL \url{https://www.oxfordmartin.ox.ac.uk/publications/structured-access-for-third-party-research-on-frontier-ai-models-investigating-researchers-model-access-requirements/}.

\bibitem[Budach et~al.(2022)Budach, Feuerpfeil, Ihde, Nathansen, Noack, Patzlaff, Naumann, and Harmouch]{budach_effects_2022}
Budach, L., Feuerpfeil, M., Ihde, N., Nathansen, A., Noack, N., Patzlaff, H., Naumann, F., and Harmouch, H.
\newblock The {Effects} of {Data} {Quality} on {Machine} {Learning} {Performance}, November 2022.
\newblock URL \url{http://arxiv.org/abs/2207.14529}.
\newblock Issue: arXiv:2207.14529 arXiv:2207.14529 [cs].

\bibitem[Buiten et~al.(2023)Buiten, de~Streel, and Peitz]{buiten_law_2023}
Buiten, M., de~Streel, A., and Peitz, M.
\newblock The law and economics of {AI} liability.
\newblock \emph{Computer Law \& Security Review}, 48:\penalty0 105794, April 2023.
\newblock ISSN 0267-3649.
\newblock \doi{10.1016/j.clsr.2023.105794}.
\newblock URL \url{https://www.sciencedirect.com/science/article/pii/S0267364923000055}.

\bibitem[Byrne et~al.(2017)Byrne, Oliner, and Sichel]{byrne2017}
Byrne, D.~M., Oliner, S.~D., and Sichel, D.~E.
\newblock How {Fast} are {Semiconductor} {Prices} {Falling}?
\newblock \emph{Review of Income and Wealth}, 64\penalty0 (3):\penalty0 679--702, April 2017.
\newblock ISSN 1475-4991.
\newblock \doi{10.1111/roiw.12308}.
\newblock URL \url{https://onlinelibrary.wiley.com/doi/abs/10.1111/roiw.12308}.
\newblock Number: 3 \_eprint: https://onlinelibrary.wiley.com/doi/pdf/10.1111/roiw.12308.

\bibitem[Carlsmith(2022)]{carlsmith_is_2022}
Carlsmith, J.
\newblock Is {Power}-{Seeking} {AI} an {Existential} {Risk}?, June 2022.
\newblock URL \url{http://arxiv.org/abs/2206.13353}.
\newblock Issue: arXiv:2206.13353 arXiv:2206.13353 [cs].

\bibitem[{Center for AI Safety}(2023)]{centerforaisafety2023}
{Center for AI Safety}.
\newblock Statement on {AI} {Risk}, 2023.
\newblock URL \url{https://www.safe.ai/statement-on-ai-risk}.

\bibitem[Character.AI(2023)]{noauthor_characterai_2023}
Character.AI.
\newblock character.ai, November 2023.
\newblock URL \url{https://beta.character.ai/help}.

\bibitem[Chen et~al.(2023)Chen, Cheng, Geng, Li, Zeng, Wang, Gong, Liu, Zeng, Dong, Tang, and Song]{chen_xtrimopglm_2023}
Chen, B., Cheng, X., Geng, Y.-a., Li, S., Zeng, X., Wang, B., Gong, J., Liu, C., Zeng, A., Dong, Y., Tang, J., and Song, L.
\newblock {xTrimoPGLM}: {Unified} {100B}-{Scale} {Pre}-trained {Transformer} for {Deciphering} the {Language} of {Protein}, July 2023.
\newblock URL \url{https://www.biorxiv.org/content/10.1101/2023.07.05.547496v1}.
\newblock Pages: 2023.07.05.547496 Section: New Results.

\bibitem[Chen et~al.(2021)Chen, Tworek, Jun, Yuan, Pinto, Kaplan, Edwards, Burda, Joseph, Brockman, Ray, Puri, Krueger, Petrov, Khlaaf, Sastry, Mishkin, Chan, Gray, Ryder, Pavlov, Power, Kaiser, Bavarian, Winter, Tillet, Such, Cummings, Plappert, Chantzis, Barnes, Herbert-Voss, Guss, Nichol, Paino, Tezak, Tang, Babuschkin, Balaji, Jain, Saunders, Hesse, Carr, Leike, Achiam, Misra, Morikawa, Radford, Knight, Brundage, Murati, Mayer, Welinder, McGrew, Amodei, McCandlish, Sutskever, and Zaremba]{chen_evaluating_2021}
Chen, M., Tworek, J., Jun, H., Yuan, Q., Pinto, H. P. d.~O., Kaplan, J., Edwards, H., Burda, Y., Joseph, N., Brockman, G., Ray, A., Puri, R., Krueger, G., Petrov, M., Khlaaf, H., Sastry, G., Mishkin, P., Chan, B., Gray, S., Ryder, N., Pavlov, M., Power, A., Kaiser, L., Bavarian, M., Winter, C., Tillet, P., Such, F.~P., Cummings, D., Plappert, M., Chantzis, F., Barnes, E., Herbert-Voss, A., Guss, W.~H., Nichol, A., Paino, A., Tezak, N., Tang, J., Babuschkin, I., Balaji, S., Jain, S., Saunders, W., Hesse, C., Carr, A.~N., Leike, J., Achiam, J., Misra, V., Morikawa, E., Radford, A., Knight, M., Brundage, M., Murati, M., Mayer, K., Welinder, P., McGrew, B., Amodei, D., McCandlish, S., Sutskever, I., and Zaremba, W.
\newblock Evaluating {Large} {Language} {Models} {Trained} on {Code}, July 2021.
\newblock URL \url{http://arxiv.org/abs/2107.03374}.
\newblock Issue: arXiv:2107.03374 arXiv:2107.03374 [cs].

\bibitem[Cottier(2023)]{cottier_trends_2023}
Cottier, B.
\newblock Trends in the {Dollar} {Training} {Cost} of {Machine} {Learning} {Systems}, January 2023.
\newblock URL \url{https://epochai.org/blog/trends-in-the-dollar-training-cost-of-machine-learning-systems}.

\bibitem[Creemers(2018)]{creemers_chinas_2018}
Creemers, R.
\newblock China's {Social} {Credit} {System}: {An} {Evolving} {Practice} of {Control}, May 2018.
\newblock URL \url{https://papers.ssrn.com/abstract=3175792}.
\newblock Issue: 3175792.

\bibitem[Demartini et~al.(2020)Demartini, Mizzaro, and Spina]{demartini2020}
Demartini, G., Mizzaro, S., and Spina, D.
\newblock Human-in-the-loop {Artiﬁcial} {Intelligence} for {Fighting} {Online} {Misinformation}: {Challenges} and {Opportunities}.
\newblock In \emph{Bulletin of the {IEEE} {Computer} {Society} {Technical} {Committee} on {Data} {Engineering}}, volume~43. IEEE Computer Society, September 2020.
\newblock URL \url{https://www.damianospina.com/publication/demartini-2020-human/demartini-2020-human.pdf}.

\bibitem[Deng et~al.(2009)Deng, Dong, Socher, Li, Li, and Fei-Fei]{deng_imagenet_2009}
Deng, J., Dong, W., Socher, R., Li, L.-J., Li, K., and Fei-Fei, L.
\newblock {ImageNet}: {A} large-scale hierarchical image database.
\newblock In \emph{2009 {IEEE} {Conference} on {Computer} {Vision} and {Pattern} {Recognition}}, pp.\  248--255, June 2009.
\newblock \doi{10.1109/CVPR.2009.5206848}.
\newblock URL \url{https://ieeexplore.ieee.org/document/5206848}.
\newblock ISSN: 1063-6919.

\bibitem[{Department for Science, Innovation {and} Technology}(2023)]{departmentforscienceinnovationandtechnology2023}
{Department for Science, Innovation {and} Technology}.
\newblock Emerging processes for frontier {AI} safety.
\newblock Policy {Paper}, UK Government, October 2023.
\newblock URL \url{https://www.gov.uk/government/publications/emerging-processes-for-frontier-ai-safety}.

\bibitem[DiEuliis et~al.(2017)DiEuliis, Carter, and Gronvall]{dieuliis_options_2017}
DiEuliis, D., Carter, S.~R., and Gronvall, G.~K.
\newblock Options for {Synthetic} {DNA} {Order} {Screening}, {Revisited}.
\newblock \emph{mSphere}, 2\penalty0 (4):\penalty0 10.1128/msphere.00319--17, August 2017.
\newblock \doi{10.1128/msphere.00319-17}.
\newblock URL \url{https://journals.asm.org/doi/full/10.1128/msphere.00319-17}.
\newblock Number: 4 Publisher: American Society for Microbiology.

\bibitem[Droppo \& Elibol(2021)Droppo and Elibol]{droppo_scaling_2021}
Droppo, J. and Elibol, O.
\newblock Scaling {Laws} for {Acoustic} {Models}, June 2021.
\newblock URL \url{http://arxiv.org/abs/2106.09488}.
\newblock Issue: arXiv:2106.09488 arXiv:2106.09488 [cs, eess].

\bibitem[Egan \& Heim(2023)Egan and Heim]{egan_oversight_2023}
Egan, J. and Heim, L.
\newblock Oversight for {Frontier} {AI} through a {Know}-{Your}-{Customer} {Scheme} for {Compute} {Providers}, October 2023.
\newblock URL \url{http://arxiv.org/abs/2310.13625}.
\newblock Issue: arXiv:2310.13625 arXiv:2310.13625 [cs].

\bibitem[{Epoch}(2022)]{epochMachineLearningData2022}
{Epoch}.
\newblock Parameter, compute and data trends in machine learning, 2022.
\newblock URL \url{https://epochai.org/data/pcd}.
\newblock tex.copyright: CC-BY.

\bibitem[{Epoch}(2023)]{epoch_ai_2023}
{Epoch}.
\newblock {AI} {Trends}, April 2023.
\newblock URL \url{https://epochai.org/trends}.

\bibitem[Erdil \& Besiroglu(2023)Erdil and Besiroglu]{erdil_algorithmic_2023}
Erdil, E. and Besiroglu, T.
\newblock Algorithmic progress in computer vision, August 2023.
\newblock URL \url{http://arxiv.org/abs/2212.05153}.
\newblock Issue: arXiv:2212.05153 arXiv:2212.05153 [cs].

\bibitem[Fitzpatrick(2021)]{fitzpatrick2021}
Fitzpatrick, A.
\newblock Why {Amazon}'s {Move} to {Drop} {Parler} {Is} a {Big} {Deal} for the {Future} of the {Internet}.
\newblock \emph{Time}, January 2021.
\newblock URL \url{https://time.com/5929888/amazon-parler-aws/}.

\bibitem[Ganguli et~al.(2022)Ganguli, Hernandez, Lovitt, DasSarma, Henighan, Jones, Joseph, Kernion, Mann, Askell, Bai, Chen, Conerly, Drain, Elhage, Showk, Fort, Hatfield-Dodds, Johnston, Kravec, Nanda, Ndousse, Olsson, Amodei, Amodei, Brown, Kaplan, McCandlish, Olah, and Clark]{ganguli_predictability_2022}
Ganguli, D., Hernandez, D., Lovitt, L., DasSarma, N., Henighan, T., Jones, A., Joseph, N., Kernion, J., Mann, B., Askell, A., Bai, Y., Chen, A., Conerly, T., Drain, D., Elhage, N., Showk, S.~E., Fort, S., Hatfield-Dodds, Z., Johnston, S., Kravec, S., Nanda, N., Ndousse, K., Olsson, C., Amodei, D., Amodei, D., Brown, T., Kaplan, J., McCandlish, S., Olah, C., and Clark, J.
\newblock Predictability and {Surprise} in {Large} {Generative} {Models}.
\newblock In \emph{2022 {ACM} {Conference} on {Fairness}, {Accountability}, and {Transparency}}, pp.\  1747--1764, June 2022.
\newblock \doi{10.1145/3531146.3533229}.
\newblock URL \url{http://arxiv.org/abs/2202.07785}.
\newblock arXiv:2202.07785 [cs].

\bibitem[Garfinkel \& Dafoe(2019)Garfinkel and Dafoe]{garfinkel_how_2019}
Garfinkel, B. and Dafoe, A.
\newblock How does the offense-defense balance scale?
\newblock \emph{Journal of Strategic Studies}, 42\penalty0 (6):\penalty0 736--763, September 2019.
\newblock ISSN 0140-2390.
\newblock \doi{10.1080/01402390.2019.1631810}.
\newblock URL \url{https://doi.org/10.1080/01402390.2019.1631810}.
\newblock Number: 6 Publisher: Routledge \_eprint: https://doi.org/10.1080/01402390.2019.1631810.

\bibitem[Google(2023)]{noauthor_bard_2023}
Google.
\newblock Bard can now connect to your {Google} apps and services, September 2023.
\newblock URL \url{https://blog.google/products/bard/google-bard-new-features-update-sept-2023/}.

\bibitem[Hendrycks et~al.(2021)Hendrycks, Burns, Basart, Zou, Mazeika, Song, and Steinhardt]{hendrycks_measuring_2021}
Hendrycks, D., Burns, C., Basart, S., Zou, A., Mazeika, M., Song, D., and Steinhardt, J.
\newblock Measuring {Massive} {Multitask} {Language} {Understanding}, January 2021.
\newblock URL \url{http://arxiv.org/abs/2009.03300}.
\newblock Issue: arXiv:2009.03300 arXiv:2009.03300 [cs].

\bibitem[Hernandez \& Brown(2020)Hernandez and Brown]{hernandez_measuring_2020}
Hernandez, D. and Brown, T.~B.
\newblock Measuring the {Algorithmic} {Efficiency} of {Neural} {Networks}, May 2020.
\newblock URL \url{http://arxiv.org/abs/2005.04305}.
\newblock Issue: arXiv:2005.04305 arXiv:2005.04305 [cs, stat].

\bibitem[Hestness et~al.(2017)Hestness, Narang, Ardalani, Diamos, Jun, Kianinejad, Patwary, Yang, and Zhou]{hestness_deep_2017}
Hestness, J., Narang, S., Ardalani, N., Diamos, G., Jun, H., Kianinejad, H., Patwary, M. M.~A., Yang, Y., and Zhou, Y.
\newblock Deep {Learning} {Scaling} is {Predictable}, {Empirically}, December 2017.
\newblock URL \url{http://arxiv.org/abs/1712.00409}.
\newblock Issue: arXiv:1712.00409 arXiv:1712.00409 [cs, stat].

\bibitem[Hobbhahn et~al.(2023)Hobbhahn, Heim, and Aydos]{hobbhahn_trends_2023}
Hobbhahn, M., Heim, L., and Aydos, G.
\newblock Trends in {Machine} {Learning} {Hardware}, November 2023.
\newblock URL \url{https://epochai.org/blog/trends-in-machine-learning-hardware}.

\bibitem[Hoffmann et~al.(2022)Hoffmann, Borgeaud, Mensch, Buchatskaya, Cai, Rutherford, Casas, Hendricks, Welbl, Clark, Hennigan, Noland, Millican, Driessche, Damoc, Guy, Osindero, Simonyan, Elsen, Rae, Vinyals, and Sifre]{hoffmann_training_2022}
Hoffmann, J., Borgeaud, S., Mensch, A., Buchatskaya, E., Cai, T., Rutherford, E., Casas, D. d.~L., Hendricks, L.~A., Welbl, J., Clark, A., Hennigan, T., Noland, E., Millican, K., Driessche, G. v.~d., Damoc, B., Guy, A., Osindero, S., Simonyan, K., Elsen, E., Rae, J.~W., Vinyals, O., and Sifre, L.
\newblock Training {Compute}-{Optimal} {Large} {Language} {Models}, March 2022.
\newblock URL \url{http://arxiv.org/abs/2203.15556}.
\newblock Issue: arXiv:2203.15556 arXiv:2203.15556 [cs].

\bibitem[Janardhan(2023)]{janardhan2023}
Janardhan, S.
\newblock Reimagining {Our} {Infrastructure} for the {AI} {Age}, May 2023.
\newblock URL \url{https://about.fb.com/news/2023/05/metas-infrastructure-for-ai/}.

\bibitem[Jouppi et~al.(2023)Jouppi, Kurian, Li, Ma, Nagarajan, Nai, Patil, Subramanian, Swing, Towles, Young, Zhou, Zhou, and Patterson]{jouppi_tpu_2023}
Jouppi, N.~P., Kurian, G., Li, S., Ma, P., Nagarajan, R., Nai, L., Patil, N., Subramanian, S., Swing, A., Towles, B., Young, C., Zhou, X., Zhou, Z., and Patterson, D.
\newblock {TPU} v4: {An} {Optically} {Reconfigurable} {Supercomputer} for {Machine} {Learning} with {Hardware} {Support} for {Embeddings}, April 2023.
\newblock URL \url{http://arxiv.org/abs/2304.01433}.
\newblock Issue: arXiv:2304.01433 arXiv:2304.01433 [cs].

\bibitem[Kaplan et~al.(2020)Kaplan, McCandlish, Henighan, Brown, Chess, Child, Gray, Radford, Wu, and Amodei]{kaplan_scaling_2020}
Kaplan, J., McCandlish, S., Henighan, T., Brown, T.~B., Chess, B., Child, R., Gray, S., Radford, A., Wu, J., and Amodei, D.
\newblock Scaling {Laws} for {Neural} {Language} {Models}, January 2020.
\newblock URL \url{http://arxiv.org/abs/2001.08361}.
\newblock Issue: arXiv:2001.08361 arXiv:2001.08361 [cs, stat].

\bibitem[Katal et~al.(2023)Katal, Dahiya, and Choudhury]{katal_energy_2023}
Katal, A., Dahiya, S., and Choudhury, T.
\newblock Energy efficiency in cloud computing data centers: a survey on software technologies.
\newblock \emph{Cluster Computing}, 26\penalty0 (3):\penalty0 1845--1875, June 2023.
\newblock ISSN 1573-7543.
\newblock \doi{10.1007/s10586-022-03713-0}.
\newblock URL \url{https://doi.org/10.1007/s10586-022-03713-0}.
\newblock Number: 3.

\bibitem[Kinniment et~al.(2023)Kinniment, Sato, Du, Goodrich, Hasin, Chan, Miles, Lin, Wijk, Burget, Ho, Barnes, and Christiano]{kinniment2023}
Kinniment, M., Sato, L. J.~K., Du, H., Goodrich, B., Hasin, M., Chan, L., Miles, L.~H., Lin, T.~R., Wijk, H., Burget, J., Ho, A., Barnes, E., and Christiano, P.
\newblock Evaluating {Language}-{Model} {Agents} on {Realistic} {Autonomous} {Tasks}.
\newblock Research paper, Alignment Research Center, 2023.

\bibitem[Knight(2023)]{knight_openais_2023}
Knight, W.
\newblock {OpenAI}’s {CEO} {Says} the {Age} of {Giant} {AI} {Models} {Is} {Already} {Over}.
\newblock \emph{Wired}, November 2023.
\newblock ISSN 1059-1028.
\newblock URL \url{https://www.wired.com/story/openai-ceo-sam-altman-the-age-of-giant-ai-models-is-already-over/}.
\newblock Section: tags.

\bibitem[Koessler \& Schuett(2023)Koessler and Schuett]{koessler_risk_2023}
Koessler, L. and Schuett, J.
\newblock Risk assessment at {AGI} companies: {A} review of popular risk assessment techniques from other safety-critical industries, July 2023.
\newblock URL \url{http://arxiv.org/abs/2307.08823}.
\newblock Issue: arXiv:2307.08823 arXiv:2307.08823 [cs].

\bibitem[Kuchinke \& Vidal(2016)Kuchinke and Vidal]{kuchinke_exclusionary_2016}
Kuchinke, B.~A. and Vidal, M.
\newblock Exclusionary strategies and the rise of winner-takes-it-all markets on the {Internet}.
\newblock \emph{Telecommunications Policy}, 40\penalty0 (6):\penalty0 582--592, June 2016.
\newblock ISSN 0308-5961.
\newblock \doi{10.1016/j.telpol.2016.02.009}.
\newblock URL \url{https://www.sciencedirect.com/science/article/pii/S0308596116000549}.
\newblock Number: 6.

\bibitem[Leike \& Sutskever(2023)Leike and Sutskever]{leike2023}
Leike, J. and Sutskever, I.
\newblock Introducing {Superalignment}, July 2023.
\newblock URL \url{https://openai.com/blog/introducing-superalignment}.

\bibitem[Li(2020)]{li2020}
Li, C.
\newblock {OpenAI}'s {GPT}-3 {Language} {Model}: {A} {Technical} {Overview}, June 2020.
\newblock URL \url{https://lambdalabs.com/blog/demystifying-gpt-3}.

\bibitem[Liang et~al.(2023)Liang, Bommasani, Lee, Tsipras, Soylu, Yasunaga, Zhang, Narayanan, Wu, Kumar, Newman, Yuan, Yan, Zhang, Cosgrove, Manning, Ré, Acosta-Navas, Hudson, Zelikman, Durmus, Ladhak, Rong, Ren, Yao, Wang, Santhanam, Orr, Zheng, Yuksekgonul, Suzgun, Kim, Guha, Chatterji, Khattab, Henderson, Huang, Chi, Xie, Santurkar, Ganguli, Hashimoto, Icard, Zhang, Chaudhary, Wang, Li, Mai, Zhang, and Koreeda]{liang_holistic_2023}
Liang, P., Bommasani, R., Lee, T., Tsipras, D., Soylu, D., Yasunaga, M., Zhang, Y., Narayanan, D., Wu, Y., Kumar, A., Newman, B., Yuan, B., Yan, B., Zhang, C., Cosgrove, C., Manning, C.~D., Ré, C., Acosta-Navas, D., Hudson, D.~A., Zelikman, E., Durmus, E., Ladhak, F., Rong, F., Ren, H., Yao, H., Wang, J., Santhanam, K., Orr, L., Zheng, L., Yuksekgonul, M., Suzgun, M., Kim, N., Guha, N., Chatterji, N., Khattab, O., Henderson, P., Huang, Q., Chi, R., Xie, S.~M., Santurkar, S., Ganguli, S., Hashimoto, T., Icard, T., Zhang, T., Chaudhary, V., Wang, W., Li, X., Mai, Y., Zhang, Y., and Koreeda, Y.
\newblock Holistic {Evaluation} of {Language} {Models}, October 2023.
\newblock URL \url{http://arxiv.org/abs/2211.09110}.
\newblock Issue: arXiv:2211.09110 arXiv:2211.09110 [cs].

\bibitem[Liu et~al.(2003)Liu, Nakashima, Sako, and Fujisawa]{liu_handwritten_2003}
Liu, C.-L., Nakashima, K., Sako, H., and Fujisawa, H.
\newblock Handwritten digit recognition: benchmarking of state-of-the-art techniques.
\newblock \emph{Pattern Recognition}, 36\penalty0 (10):\penalty0 2271--2285, October 2003.
\newblock ISSN 0031-3203.
\newblock \doi{10.1016/S0031-3203(03)00085-2}.
\newblock URL \url{https://www.sciencedirect.com/science/article/pii/S0031320303000852}.
\newblock Number: 10.

\bibitem[Lohn \& Jackson(2022)Lohn and Jackson]{lohn2022}
Lohn, A. and Jackson, K.
\newblock Will {AI} {Make} {Cyber} {Swords} or {Shields}?, August 2022.
\newblock URL \url{https://cset.georgetown.edu/publication/will-ai-make-cyber-swords-or-shields/}.

\bibitem[Lohn \& Musser(2022)Lohn and Musser]{lohn2022a}
Lohn, A. and Musser, M.
\newblock {AI} and {Compute}: {How} {Much} {Longer} {Can} {Computing} {Power} {Drive} {Artificial} {Intelligence} {Progress}?
\newblock Technical report, Center for Security and Emerging Technology, January 2022.
\newblock URL \url{https://cset.georgetown.edu/publication/ai-and-compute/}.

\bibitem[McGregor(2021)]{mcgregor_preventing_2021}
McGregor, S.
\newblock Preventing {Repeated} {Real} {World} {AI} {Failures} by {Cataloging} {Incidents}: {The} {AI} {Incident} {Database}.
\newblock \emph{Proceedings of the AAAI Conference on Artificial Intelligence}, 35\penalty0 (17):\penalty0 15458--15463, May 2021.
\newblock ISSN 2374-3468.
\newblock \doi{10.1609/aaai.v35i17.17817}.
\newblock URL \url{https://ojs.aaai.org/index.php/AAAI/article/view/17817}.
\newblock Number: 17.

\bibitem[Meta(2023)]{meta2023}
Meta.
\newblock Introducing {Llama} 2, 2023.
\newblock URL \url{https://ai.meta.com/blog/llama-2/}.

\bibitem[Mistral(2023)]{aiMixtralExperts2023}
Mistral.
\newblock Mixtral of experts.
\newblock https://mistral.ai/news/mixtral-of-experts/, December 2023.

\bibitem[Mulani \& Whittlestone(2023)Mulani and Whittlestone]{mulani2023}
Mulani, N. and Whittlestone, J.
\newblock Proposing a {Foundation} {Model} {Information}-{Sharing} {Regime} for the {UK} {\textbar} {GovAI} {Blog}.
\newblock Research {Post}, Centre for the Governance of AI, June 2023.
\newblock URL \url{https://www.governance.ai/post/proposing-a-foundation-model-information-sharing-regime-for-the-uk#}.

\bibitem[Nelson \& Rose(2023)Nelson and Rose]{nelson2023}
Nelson, C. and Rose, S.
\newblock Report launch: examining risks at the intersection of {AI} and bio.
\newblock Research {Report}, Centre for Long-Term Resilience, October 2023.
\newblock URL \url{https://www.longtermresilience.org/post/report-launch-examining-risks-at-the-intersection-of-ai-and-bio}.

\bibitem[Nevo et~al.(2023)Nevo, Lahav, Karpur, Alstott, and Matheny]{nevo_securing_2023}
Nevo, S., Lahav, D., Karpur, A., Alstott, J., and Matheny, J.
\newblock Securing {Artificial} {Intelligence} {Model} {Weights}: {Interim} {Report}.
\newblock Technical report, RAND Corporation, October 2023.
\newblock URL \url{https://www.rand.org/pubs/working_papers/WRA2849-1.html}.

\bibitem[Ngo et~al.(2023)Ngo, Chan, and Mindermann]{ngo_alignment_2023}
Ngo, R., Chan, L., and Mindermann, S.
\newblock The alignment problem from a deep learning perspective, September 2023.
\newblock URL \url{http://arxiv.org/abs/2209.00626}.
\newblock Issue: arXiv:2209.00626 arXiv:2209.00626 [cs].

\bibitem[NIST(2023)]{noauthor_face_2023}
NIST.
\newblock Face {Recognition} {Technology} {Evaluation} ({FRTE}) 1:1 {Verification}, November 2023.
\newblock URL \url{https://pages.nist.gov/frvt/html/frvt11.html}.

\bibitem[Novelli et~al.(2023)Novelli, Casolari, Rotolo, Taddeo, and Floridi]{novelli_taking_2023}
Novelli, C., Casolari, F., Rotolo, A., Taddeo, M., and Floridi, L.
\newblock Taking {AI} risks seriously: a new assessment model for the {AI} {Act}.
\newblock \emph{AI \& SOCIETY}, July 2023.
\newblock ISSN 1435-5655.
\newblock \doi{10.1007/s00146-023-01723-z}.
\newblock URL \url{https://doi.org/10.1007/s00146-023-01723-z}.

\bibitem[O'Brien et~al.(2023)O'Brien, Ee, and Williams]{obrien2023a}
O'Brien, J., Ee, S., and Williams, Z.
\newblock Deployment corrections: {An} incident response framework for frontier {AI} models.
\newblock Research {Report}, Institute for AI Policy and Strategy, September 2023.
\newblock URL \url{https://www.iaps.ai/research/deployment-corrections}.

\bibitem[OpenAI(2023)]{openai2023b}
OpenAI.
\newblock {GPT}-4, March 2023.
\newblock URL \url{https://openai.com/research/gpt-4}.

\bibitem[{OpenAI}(2023)]{openai_gpt-4_2023}
{OpenAI}.
\newblock {GPT}-4 {Technical} {Report}, March 2023.
\newblock URL \url{http://arxiv.org/abs/2303.08774}.
\newblock Issue: arXiv:2303.08774 arXiv:2303.08774 [cs].

\bibitem[{Our World in Data}(2023)]{noauthor_computation_2023}
{Our World in Data}.
\newblock Computation used to train notable {AI} systems, by affiliation of researchers, November 2023.
\newblock URL \url{https://ourworldindata.org/grapher/artificial-intelligence-training-computation-by-researcher-affiliation}.

\bibitem[{Papers with Code}(2023)]{noauthor_papers_2023}
{Papers with Code}.
\newblock Papers with {Code} - {MMLU} {Benchmark} ({Multi}-task {Language} {Understanding}), November 2023.
\newblock URL \url{https://paperswithcode.com/sota/multi-task-language-understanding-on-mmlu}.

\bibitem[Patel(2023)]{patel2023}
Patel, D.
\newblock Dario {Amodei} ({Anthropic} {CEO}) - {Scaling}, {Alignment}, \& {AI} {Progress}, August 2023.
\newblock URL \url{https://www.dwarkeshpatel.com/p/dario-amodei}.
\newblock Dwarkesh Podcast.

\bibitem[Peng et~al.(2023)Peng, Wu, Allard, Kilpatrick, and Heidel]{peng2023}
Peng, A., Wu, M., Allard, J., Kilpatrick, L., and Heidel, S.
\newblock {GPT}-3.5 {Turbo} fine-tuning and {API} updates, August 2023.
\newblock URL \url{https://openai.com/blog/gpt-3-5-turbo-fine-tuning-and-api-updates}.

\bibitem[Radford et~al.(2019)Radford, Wu, Child, Luan, Amodei, and Sutskever]{radford2019}
Radford, A., Wu, J., Child, R., Luan, D., Amodei, D., and Sutskever, I.
\newblock Language {Models} are {Unsupervised} {Multitask} {Learners}, 2019.
\newblock URL \url{https://d4mucfpksywv.cloudfront.net/better-language-models/language_models_are_unsupervised_multitask_learners.pdf}.

\bibitem[Richter(2023)]{richter2023}
Richter, F.
\newblock Amazon {Maintains} {Lead} in the {Cloud} {Market}, August 2023.
\newblock URL \url{https://www.statista.com/chart/18819/worldwide-market-share-of-leading-cloud-infrastructure-service-providers}.

\bibitem[Roser(2023)]{roser_artificial_2023}
Roser, M.
\newblock Artificial intelligence has advanced despite having few resources dedicated to its development – now investments have increased substantially.
\newblock \emph{Our World in Data}, October 2023.
\newblock URL \url{https://ourworldindata.org/ai-investments}.

\bibitem[Roser et~al.(2023)Roser, Ritchie, and Mathieu]{roser2023a}
Roser, M., Ritchie, H., and Mathieu, E.
\newblock What is {Moore}’s {Law}?, March 2023.
\newblock URL \url{https://ourworldindata.org/moores-law}.

\bibitem[Russell(2019)]{russell_human_2019}
Russell, S.
\newblock \emph{Human {Compatible}: {Artificial} {Intelligence} and the {Problem} of {Control}}.
\newblock Penguin, October 2019.
\newblock ISBN 978-0-525-55862-0.

\bibitem[Sandbrink(2023)]{sandbrink_artificial_2023}
Sandbrink, J.~B.
\newblock Artificial intelligence and biological misuse: {Differentiating} risks of language models and biological design tools, October 2023.
\newblock URL \url{http://arxiv.org/abs/2306.13952}.
\newblock Issue: arXiv:2306.13952 arXiv:2306.13952 [cs].

\bibitem[Sedova et~al.(2021)Sedova, McNeill, Johnson, Joshi, and Wulkan]{sedova2021}
Sedova, K., McNeill, C., Johnson, A., Joshi, A., and Wulkan, I.
\newblock {AI} and the {Future} of {Disinformation} {Campaigns}.
\newblock {CSET} {Policy} {Brief}, Centre for Security and Emerging Technology, December 2021.
\newblock URL \url{https://cset.georgetown.edu/wp-content/uploads/CSET-AI-and-the-Future-of-Disinformation-Campaigns-Part-2.pdf}.

\bibitem[Seger et~al.(2023)Seger, Dreksler, Moulange, Dardaman, Schuett, Wei, Winter, Arnold, Ó~hÉigeartaigh, Korinek, Anderljung, Bucknall, Chan, Stafford, Koessler, Ovadya, Garfinkel, Bluemke, Aird, Levermore, Hazell, and Gupta]{seger2023}
Seger, E., Dreksler, N., Moulange, R., Dardaman, E., Schuett, J., Wei, K., Winter, C., Arnold, M., Ó~hÉigeartaigh, S., Korinek, A., Anderljung, M., Bucknall, B., Chan, A., Stafford, E., Koessler, L., Ovadya, A., Garfinkel, B., Bluemke, E., Aird, M., Levermore, P., Hazell, J., and Gupta, A.
\newblock Open-{Sourcing} {Highly} {Capable} {Foundation} {Models}.
\newblock Research paper, Centre for the Governance of AI, September 2023.
\newblock URL \url{https://www.governance.ai/research-paper/open-sourcing-highly-capable-foundation-models}.

\bibitem[Senior et~al.(2020)Senior, Evans, Jumper, Kirkpatrick, Sifre, Green, Qin, Žídek, Nelson, Bridgland, Penedones, Petersen, Simonyan, Crossan, Kohli, Jones, Silver, Kavukcuoglu, and Hassabis]{senior_improved_2020}
Senior, A.~W., Evans, R., Jumper, J., Kirkpatrick, J., Sifre, L., Green, T., Qin, C., Žídek, A., Nelson, A. W.~R., Bridgland, A., Penedones, H., Petersen, S., Simonyan, K., Crossan, S., Kohli, P., Jones, D.~T., Silver, D., Kavukcuoglu, K., and Hassabis, D.
\newblock Improved protein structure prediction using potentials from deep learning.
\newblock \emph{Nature}, 577\penalty0 (7792):\penalty0 706--710, January 2020.
\newblock ISSN 1476-4687.
\newblock \doi{10.1038/s41586-019-1923-7}.
\newblock URL \url{https://www.nature.com/articles/s41586-019-1923-7}.
\newblock Number: 7792 Publisher: Nature Publishing Group.

\bibitem[Sevilla et~al.(2022)Sevilla, Heim, Ho, Besiroglu, Hobbhahn, and Villalobos]{sevilla_compute_2022}
Sevilla, J., Heim, L., Ho, A., Besiroglu, T., Hobbhahn, M., and Villalobos, P.
\newblock Compute {Trends} {Across} {Three} {Eras} of {Machine} {Learning}, March 2022.
\newblock URL \url{http://arxiv.org/abs/2202.05924}.
\newblock Issue: arXiv:2202.05924 arXiv:2202.05924 [cs].

\bibitem[Shevlane(2022)]{shevlane_structured_2022}
Shevlane, T.
\newblock Structured access: an emerging paradigm for safe {AI} deployment, April 2022.
\newblock URL \url{http://arxiv.org/abs/2201.05159}.
\newblock Issue: arXiv:2201.05159 arXiv:2201.05159 [cs].

\bibitem[Shevlane et~al.(2023)Shevlane, Farquhar, Garfinkel, Phuong, Whittlestone, Leung, Kokotajlo, Marchal, Anderljung, Kolt, Ho, Siddarth, Avin, Hawkins, Kim, Gabriel, Bolina, Clark, Bengio, Christiano, and Dafoe]{shevlane_model_2023}
Shevlane, T., Farquhar, S., Garfinkel, B., Phuong, M., Whittlestone, J., Leung, J., Kokotajlo, D., Marchal, N., Anderljung, M., Kolt, N., Ho, L., Siddarth, D., Avin, S., Hawkins, W., Kim, B., Gabriel, I., Bolina, V., Clark, J., Bengio, Y., Christiano, P., and Dafoe, A.
\newblock Model evaluation for extreme risks, September 2023.
\newblock URL \url{http://arxiv.org/abs/2305.15324}.
\newblock Issue: arXiv:2305.15324 arXiv:2305.15324 [cs].

\bibitem[Solaiman(2023)]{solaiman_gradient_2023}
Solaiman, I.
\newblock The {Gradient} of {Generative} {AI} {Release}: {Methods} and {Considerations}, February 2023.
\newblock URL \url{http://arxiv.org/abs/2302.04844}.
\newblock Issue: arXiv:2302.04844 arXiv:2302.04844 [cs].

\bibitem[{The White House}(2023)]{house_executive_2023}
{The White House}.
\newblock Executive {Order} on the {Safe}, {Secure}, and {Trustworthy} {Development} and {Use} of {Artificial} {Intelligence}, October 2023.
\newblock URL \url{https://www.whitehouse.gov/briefing-room/presidential-actions/2023/10/30/executive-order-on-the-safe-secure-and-trustworthy-development-and-use-of-artificial-intelligence/}.

\bibitem[Tucker et~al.(2020)Tucker, Anderljung, and Dafoe]{tucker_social_2020}
Tucker, A.~D., Anderljung, M., and Dafoe, A.
\newblock Social and {Governance} {Implications} of {Improved} {Data} {Efficiency}.
\newblock In \emph{Proceedings of the {AAAI}/{ACM} {Conference} on {AI}, {Ethics}, and {Society}}, pp.\  378--384, February 2020.
\newblock \doi{10.1145/3375627.3375863}.
\newblock URL \url{http://arxiv.org/abs/2001.05068}.
\newblock arXiv:2001.05068 [cs].

\bibitem[Urbina et~al.(2022)Urbina, Lentzos, Invernizzi, and Ekins]{urbina_dual_2022}
Urbina, F., Lentzos, F., Invernizzi, C., and Ekins, S.
\newblock Dual use of artificial-intelligence-powered drug discovery.
\newblock \emph{Nature Machine Intelligence}, 4\penalty0 (3):\penalty0 189--191, March 2022.
\newblock ISSN 2522-5839.
\newblock \doi{10.1038/s42256-022-00465-9}.
\newblock URL \url{https://www.nature.com/articles/s42256-022-00465-9}.
\newblock Number: 3 Publisher: Nature Publishing Group.

\bibitem[Venigalla \& Li(2022)Venigalla and Li]{venigalla2022}
Venigalla, A. and Li, L.
\newblock Mosaic {LLMs} ({Part} 2): {GPT}-3 quality for {\textless}\$500k, September 2022.
\newblock URL \url{https://www.mosaicml.com/blog/gpt-3-quality-for-500k}.

\bibitem[Villalobos(2023)]{villalobos_scaling_2023}
Villalobos, P.
\newblock Scaling {Laws} {Literature} {Review}, January 2023.
\newblock URL \url{https://epochai.org/blog/scaling-laws-literature-review}.

\bibitem[Villalobos et~al.(2022)Villalobos, Sevilla, Heim, Besiroglu, Hobbhahn, and Ho]{villalobos_will_2022}
Villalobos, P., Sevilla, J., Heim, L., Besiroglu, T., Hobbhahn, M., and Ho, A.
\newblock Will we run out of data? {An} analysis of the limits of scaling datasets in {Machine} {Learning}, October 2022.
\newblock URL \url{http://arxiv.org/abs/2211.04325}.
\newblock Issue: arXiv:2211.04325 arXiv:2211.04325 [cs].

\bibitem[Weidinger et~al.(2021)Weidinger, Mellor, Rauh, Griffin, Uesato, Huang, Cheng, Glaese, Balle, Kasirzadeh, Kenton, Brown, Hawkins, Stepleton, Biles, Birhane, Haas, Rimell, Hendricks, Isaac, Legassick, Irving, and Gabriel]{weidinger_ethical_2021}
Weidinger, L., Mellor, J., Rauh, M., Griffin, C., Uesato, J., Huang, P.-S., Cheng, M., Glaese, M., Balle, B., Kasirzadeh, A., Kenton, Z., Brown, S., Hawkins, W., Stepleton, T., Biles, C., Birhane, A., Haas, J., Rimell, L., Hendricks, L.~A., Isaac, W., Legassick, S., Irving, G., and Gabriel, I.
\newblock Ethical and social risks of harm from {Language} {Models}, December 2021.
\newblock URL \url{http://arxiv.org/abs/2112.04359}.
\newblock Issue: arXiv:2112.04359 arXiv:2112.04359 [cs].

\bibitem[Wu et~al.(2023)Wu, Irsoy, Lu, Dabravolski, Dredze, Gehrmann, Kambadur, Rosenberg, and Mann]{wu_bloomberggpt_2023}
Wu, S., Irsoy, O., Lu, S., Dabravolski, V., Dredze, M., Gehrmann, S., Kambadur, P., Rosenberg, D., and Mann, G.
\newblock {BloombergGPT}: {A} {Large} {Language} {Model} for {Finance}, May 2023.
\newblock URL \url{http://arxiv.org/abs/2303.17564}.
\newblock Issue: arXiv:2303.17564 arXiv:2303.17564 [cs, q-fin] version: 2.

\bibitem[Yu et~al.(2022)Yu, Xu, Koh, Luong, Baid, Wang, Vasudevan, Ku, Yang, Ayan, Hutchinson, Han, Parekh, Li, Zhang, Baldridge, and Wu]{yu_scaling_2022}
Yu, J., Xu, Y., Koh, J.~Y., Luong, T., Baid, G., Wang, Z., Vasudevan, V., Ku, A., Yang, Y., Ayan, B.~K., Hutchinson, B., Han, W., Parekh, Z., Li, X., Zhang, H., Baldridge, J., and Wu, Y.
\newblock Scaling {Autoregressive} {Models} for {Content}-{Rich} {Text}-to-{Image} {Generation}, June 2022.
\newblock URL \url{http://arxiv.org/abs/2206.10789}.
\newblock Issue: arXiv:2206.10789 arXiv:2206.10789 [cs].

\bibitem[Zhang et~al.(2022)Zhang, Maslej, Brynjolfsson, Etchemendy, Lyons, Manyika, Ngo, Niebles, Sellitto, Sakhaee, Shoham, Clark, and Perrault]{zhang2022}
Zhang, D., Maslej, N., Brynjolfsson, E., Etchemendy, J., Lyons, T., Manyika, J., Ngo, H., Niebles, J.~C., Sellitto, M., Sakhaee, E., Shoham, Y., Clark, J., and Perrault, R.
\newblock The {AI} {Index} 2022 {Annual} {Report}.
\newblock Technical report, AI Index Steering Committee, Stanford Institute for Human-Centered AI, Stanford University, March 2022.
\newblock URL \url{https://aiindex.stanford.edu/wp-content/uploads/2022/03/2022-AI-Index-Report_Master.pdf}.

\end{thebibliography}
\bibliographystyle{icml2024}

\newpage
\appendix




\end{document}